\documentclass[11pt]{article}
\pdfoutput=1
\usepackage{dcolumn}
\usepackage{bm}
\usepackage{graphicx}
\usepackage{amssymb,amsmath}
\usepackage{multirow}
\usepackage{cite,color,url}
\usepackage[colorlinks=true
,urlcolor=blue
,anchorcolor=blue
,citecolor=blue
,filecolor=blue
,linkcolor=blue
,menucolor=blue
,linktocpage=true
,pdfproducer=medialab
,pdfa=true3
]{hyperref}

\usepackage{slashed}
\usepackage{epsfig,psfrag,rotating,soul}
\usepackage{rotfloat}
\usepackage[font={small}]{caption}
\usepackage{colortbl}
\usepackage{xcolor}
\definecolor{darkspringgreen}{rgb}{0.09, 0.45, 0.27}

\usepackage{extarrows} 

\evensidemargin \oddsidemargin
\allowdisplaybreaks

\let\OLDthebibliography\thebibliography
\renewcommand\thebibliography[1]{
  \OLDthebibliography{#1}
  \setlength{\parskip}{0pt}
  \setlength{\itemsep}{0pt plus 0.3ex}
}

\definecolor{mygray}{gray}{0.85} 
\definecolor{myblue}{cmyk}{0.65, 0.37, 0.0, 0.19}

\begin{document}
\thispagestyle{empty}

\def\thefootnote{\fnsymbol{footnote}}

\vspace*{1cm}

\begin{center}

\begin{Large}
\textbf{\textsc{Probing triple-gauge couplings in anomalous gauge theories\\ at hadron and lepton colliders}}
\end{Large}

\vspace{1cm}

{\sc
Anibal~D.~Medina$^{1}$%
\footnote{{\tt \href{mailto:anibal.medina@fisica.unlp.edu.ar}{anibal.medina@fisica.unlp.edu.ar}}}%
, Nicol\'as~I.~Mileo$^{1}$%
\footnote{{\tt \href{mailto:mileo@fisica.unlp.edu.ar}{mileo@fisica.unlp.edu.ar}}}%
, Alejandro~Szynkman$^{1}$%
\footnote{{\tt \href{mailto:szynkman@fisica.unlp.edu.ar}{szynkman@fisica.unlp.edu.ar}}}%
,  Santiago~A.~Tanco$^{1}$%
\footnote{{\tt \href{mailto:santiago.tanco@fisica.unlp.edu.ar}{santiago.tanco@fisica.unlp.edu.ar}}}%
, Carlos~E.~M.~Wagner$^{2, 3, 4}$%
\footnote{{\tt \href{mailto:cwagner@anl.gov}{cwagner@anl.gov}}}%
and Gabriel~Zapata$^{5}$%
\footnote{{\tt \href{mailto:gabriel.zapata@pucp.edu.pe}{gabriel.zapata@pucp.edu.pe}}}%

}

\vspace*{.7cm}

{\sl

$^1$IFLP, CONICET - Dpto. de F\'{\i}sica, Universidad Nacional de La Plata,  C.C. 67, 1900 La Plata,  Argentina

\vspace*{0.1cm}

$^2$HEP Division, Argonne National Laboratory, 9700 Cass Ave., Argonne, IL 60439, USA

\vspace*{0.1cm}

$^3$Enrico Fermi Institute, Physics Department, University of Chicago, Chicago, IL 60637, USA

\vspace*{0.1cm}

$^4$Kavli Institute for Cosmological Physics, University of Chicago, Chicago, IL 60637, USA

\vspace*{0.1cm}

$^5$Secci\'on F\'isica, Departamento de Ciencias, Pontificia Universidad Cat\'olica del Per\'u, Apartado 1761, Lima, Per\'u

}

\end{center}

\vspace{0.1cm}

\begin{abstract}
\noindent

Gauge anomalous quantum field theories are inconsistent as full UV theories since they lead to the breaking of Lorentz invariance or Unitarity, as well as non-renormalizability. It is well known, however, that they can be interpreted as effective field theories (EFT) with a cut-off. The latter cannot be made arbitrarily large and it is related to the energy scale at which additional fermions with suitable gauge charges enter, rendering the full model anomaly-free. A nondecoupling effect that remains in the EFT is the appearance of anomalous loop-induced triple-gauge couplings, encapsulating information from the full UV theory. In this work we take as an example an Abelian gauge symmetry $U(1)'_\mu$ under which $2^{nd}$-generation leptons are axially charged, leading to an EFT that consists of the Standard Model (SM) with an additional massive $Z'$ gauge boson. As a consequence, there are triple gauge couplings involving the $Z'$ and Electroweak SM gauge bosons via mixed gauge anomalies. We study the possibility of probing these loop suppressed anomalous couplings at hadron and lepton colliders, with $Z'$-lepton couplings allowed by current experimental bounds, finding that due to the large SM backgrounds and small signal, the HL-LHC is incapable of this task. The 100 TeV $pp$ collider at $\mathcal{L}=20~\mathrm   {ab}^{-1}$ on the other hand could probe anomalous couplings for  $m_{Z'}\in[150,800]~\mathrm{GeV}$ and obtain discovery significances for $m_{Z'}\in[230,330]~\mathrm{GeV}$. Lepton colliders are also well suited for probing these anomalous couplings. In particular we show that a muon collider running at the $Z'$-resonance and an electron-positron collider such as CLIC with $\sqrt{s}=3$ TeV can be complimentary in probing the anomalous couplings for $m_{Z'}\in[100,700]$ GeV, with CLIC sensitive to discovery for $m_{Z'}\in[125,225]\; {\rm GeV}$.

\end{abstract}

\def\thefootnote{\arabic{footnote}}
\setcounter{page}{0}
\setcounter{footnote}{0}

\newpage

\section{Introduction}
\label{intro}

Interactions among three electroweak (EW) gauge bosons dominated by one-loop triangle Feynman diagrams have been actively searched for at LEP~\cite{DELPHI:2007gzg}, Tevatron~\cite{D0:2008swt,D0:2011pfu,CDF:2011rqn} and the LHC~\cite{ATLAS:2011ahl,ATLAS:2011tia,Martelli:2012np}. 
In the Standard Model (SM) and at energies $\sqrt{s}\gtrsim 2 m_Z$, with $m_Z$ the $Z$ gauge boson mass, the top quark contribution in the loop dominates due to its much larger mass in comparison with the rest of the fermionic matter content. These vertices peak at $\sqrt{s}\gtrsim 2 m_t$ and fall as $1/s$ at large energies~\footnote{ The $Z^{*}\gamma\gamma$ falls as $1/s^2$ at $\sqrt{s}\gg m_t$.} due to the anomaly free nature of the SM. 
Even at their peak values, considering the production of an off-shell gauge boson and its subsequent decay via the loop-induced vertices to two gauge bosons, neither of the past nor current colliders at the their largest projected luminosities are able to probe these triple gauge couplings~\cite{Dedes:2012me,Hernandez-Juarez:2021mhi}.

A possible window into probing triple gauge couplings opens up when considering quantum gauge anomalous Abelian extensions of the SM as low-energy effective theories with a natural cut-off. This happens, for example, in the case that part of the SM matter content is at least axially charged under the new Abelian gauge group.  In that case, the mixed gauge anomalies provided by the loop-induced vertices of the new Abelian gauge boson to two EW SM gauge bosons may, under certain kinematical conditions, allow for an enhancement in the production of gauge bosons via triple gauge couplings.  
It is well known that theories with gauge anomalies are sick, since in fact gauge symmetry is explicitly broken at the quantum level as shown by the violations of the Ward identities, ultimately leading to the breaking of Unitarity or Lorentz invariance and nonrenormalizability~\cite{Preskill:1990fr}. Part of the sickness can be solved considering that the Abelian gauge group is spontaneously broken by the vacuum expectation value of a Higgs $\langle \Phi \rangle$, making the associated $Z'$ gauge boson massive~\footnote{Abelian gauge groups allow the possibility of keeping the $Z'$ in the low energy effective theory via a small gauge coupling, while for the non-Abelian case, the $Z'$ mass is fixed by the group structure and of the order of  $\langle \Phi \rangle$.}. The issue of renormalizability  however remains and implies that the theory can only be regarded as an effective theory with a cut-off $\Lambda$ that cannot be made arbitrarily large without suffering a loss of calculability. In fact, at energies of the order of $\Lambda$, new physics must enter into the theory rendering the full model anomaly free. This new physics can usually be interpreted in the form of additional fermions, sometimes referred to as spectators, whose charges under the Abelian gauge group are such that they lead to a cancellation of the anomalous terms from the axially charged SM matter content. At low energies, in the effective theory where the spectator fermions have been integrated out, their non-decoupling effect remains in the form of a Wess-Zumino-Witten (WZW) term in the effective action~\cite{Dror:2017nsg} and a freedom exists in the coefficients that enter the WZW term related to the regularization scheme adopted~\footnote{In fact there is a one-to-one correspondence between the WZW coefficients and a momentum shift in the loop.}. Given that we do not want the anomaly to affect the SM gauge groups, we adopt what is known as the covariant regularization scheme, which in conjunction with the Wess-Zumino consistency condition, fixes the value of the WZW coefficients~\cite{Ismail:2017ulg, Ismail:2017fgq }. In this way, all gauge anomalous effects are transferred to the $Z'$ modified Ward identity.  We show that it is precisely this anomaly-induced triple gauge boson vertex, via the $Z'$ longitudinal polarization, that may lead to an enhancement in the searched signals. 
 
In order to consider an explicit example, we study the case of a muonic $U(1)'_\mu$, under which muons and their corresponding neutrinos are axially charged. This scenario was part of a previous work~\cite{Medina:2021ram}, in which dark matter phenomenology played a major role. For our current study, however, to maximize the signal we do not consider a DM particle, and any missing energy signal in our collider studies comes from neutrinos. Focusing on the anomalous triple gauge boson vertices $Z'ZZ$, $Z'\gamma\gamma$ and $Z'Z\gamma$ and implementing them in \texttt{MadGraph}~\cite{Alwall:2014hca}, we study the ability of different colliders in probing these anomalous triple gauge couplings taking into account, as a first approximation, only the irreducible SM backgrounds. 
We look first at hadron colliders, focusing on the High-Luminosity LHC (HL-LHC) at $\sqrt{s}=14$ TeV and the 100 TeV hadron collider and study their ability to test the anomalous triple gauge couplings. 
Afterwards, we move into leptonic colliders, considering a muon collider running at $\sqrt{s}=m_{Z^{\prime}}$ and the  $e^{+}e^{-}$  collider CLIC at $\sqrt{s}=3$~TeV.

This paper is organized as follows. In Sect.~\ref{sec:model} we do a brief introduction into the specific effective model considered and the anomalous triple gauge vertices that we focus on for collider studies. In Sect.~\ref{sec:hadroncoll} we study the ability of hadron colliders, the HL-LHC and the 100 TeV, in probing anomalous triple gauge couplings. In Sect.~\ref{sec:leptoncoll} we show that lepton colliders, $\mu^{+}\mu^{-}$ and $e^{+}e^{-}$, have a much better chance in probing these couplings and finally in Sect.~\ref{sec:conclusion} we give our conclusions.

\section{Anomalous $U(1)'_\mu$ effective theory and event simulation}
\label{sec:model}

We consider a model where only the leptons of the second generation are charged under a gauge symmetry $U(1)'_\mu$ and the new interaction for the muon in the mass basis is axial~\cite{Medina:2021ram}. All remaining SM fields, including the SM Higgs, are neutral under $U(1)'_\mu$~\footnote{Recent studies with muon-philic models have been published in~\cite{Kawamura:2019rth,Abdughani:2021oit,Perelstein:2020suc}.}. We assume that the $U(1)'_\mu$ symmetry is broken by some scalar field with a non-zero vacuum expectation value and we define at some smaller scale the effective theory of the SM gauge field and matter content along with a massive gauge vector boson $Z'$. The EFT can be trusted up to an energy scale of order $\Lambda\lesssim 64 \pi^3 m_{Z'}/(3 g_{\mu}g^2_{SM})$~\cite{Preskill:1990fr,Ismail:2017fgq}, which for the values we consider is of the order of 800 TeV. Within this framework, the Higgs field that triggers the spontaneous breaking of the $U(1)'_\mu$ is supposed to have a sufficiently large mass to be integrated out from the effective theory~\footnote{A new Higgs might be not required and the $Z'$ mass could also be produced by the Stückelberg mechanism.},

\begin{equation}
\label{eq1}
    \mathcal{L}=\mathcal{L}_{SM}-\frac{1}{4}Z'_{\mu\nu}Z'^{\mu\nu} + \frac{1}{2}m^2_{Z'} Z'_\rho Z'^\rho+ g_\mu \bar{\mu}\gamma^\rho\gamma^5\mu Z'_\rho - g_\mu \bar{\nu}_{\mu L}\gamma^\rho\nu_{\mu L} Z'_\rho \; , \vspace*{0.5cm}
\end{equation}
where $Z'_{\mu\nu}=\partial_{\mu}Z'_{\nu}-\partial_{\nu}Z'_{\mu}$ stands for the $Z'$-field strength, $g_\mu=Q_{\mu}g'$ is the coupling strength for the interactions of the muon and neutrino which have charge $Q_{\mu}$ under $U(1)'_\mu$, $g'$ denotes the $U(1)'_\mu$ coupling, and $m_{Z'}$ is the mass of the $Z'$ gauge boson. The $Z'-\nu_\mu$ interaction is set to maintain the EW symmetry. No tree-level kinetic mixing term among the SM EW gauge bosons and the $Z'$ is assumed.

From the Lagrangian in Eq.~(\ref{eq1}) we derive the partial decay widths of the $Z'$ boson at leading order
\begin{eqnarray}
\label{eq2}
\Gamma(Z'\to \mu^+\mu^-)&=& \frac{g^2_{\mu}m_{Z'}}{12\pi}(1-4z_{\mu})^{3/2},\\[1mm]
\label{eq3}
\Gamma(Z'\to \nu_{\mu}\bar{\nu}_{\mu})&=& \frac{g^2_{\mu}m_{Z'}}{24\pi},
\end{eqnarray}
where $z_{\mu}=m^2_\mu/m^2_{Z'}$. If the channels in Eqs.~(\ref{eq2})-(\ref{eq3}) saturate the total width, the branching ratio into muonic neutrinos is $\mathrm{BR}(Z'\to\nu_{\mu}\bar{\nu}_{\mu})= \frac{1}{3}$ up to corrections $\mathcal{O}(z_{\mu})$. The ratio between the total decay width, $\Gamma_{Z'}$, and the $Z'$ mass in terms of the coupling $g_{\mu}$ is
\begin{equation}
\label{eq6}
\frac{\Gamma_{Z'}}{m_{Z'}}=\frac{g^2_{\mu}}{8\pi}.
\end{equation}

Notice that the muon mass cannot be generated as usual via EW symmetry breaking within this model. Being the muon both charged electromagnetically and under the $U(1)'_{\mu}$ gauge symmetry, the ordinary muon Yukawa interaction would explicitly break $U(1)'_{\mu}$. It is possible to recover this interaction at low energy from a higher-dimensional operator which combines the ordinary Yukawa interaction and a SM singlet Higgs field that induces the spontaneous breaking of the $U(1)'_{\mu}$ symmetry. A discussion about this issue is presented in Sect.~2 of reference~\cite{Medina:2021ram}. 

It is crucial to note here that this effective theory contains gauge anomalies which result naively in the breakdown of gauge invariance and/or Unitarity with the consequent appearance of inconsistencies at the quantum level. The source of these anomalies is traced back to the vector-axial nature of the leptonic current coupled to the new gauge boson and the fact that it is only the second generation of leptons which is charged under the new gauge symmetry. In the effective theory the gauge anomalies generate in particular anomalous triple gauge boson couplings involving the $U(1)'_{\mu}$ and SM gauge bosons, known as mixed anomalies. Since gauge anomalies must be certainly absent in the full UV theory, new fermions are required to cancel all the anomalies present at low energies~\footnote{See Section 5.1 of reference~\cite{Medina:2021ram} for details regarding the anomalies cancellation.}. These new fermions in turn affect the effective theory through their effects on the triple gauge-boson couplings via the WZW term.

\begin{figure}
    \centering
    \includegraphics[width=0.3\linewidth]{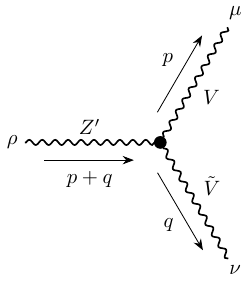}
    \caption{Triple gauge boson coupling between $Z'$ and two EW bosons $V,\tilde{V}$. Momenta labels and indices match Eq.~\eqref{rosenberg}.}
    \label{fig:tribosondiag}
\end{figure}

The anomalous $Z'V\tilde{V}$ triple gauge couplings between $Z'$ and two EW gauge bosons, see Fig~\ref{fig:tribosondiag}, have the general form consistent with Lorentz symmetry given by the Rosenberg parametrization~\cite{Rosenberg:1962pp},
\begin{eqnarray}
\begin{aligned}
A_{\rho \mu \nu}^{Z' V \tilde{V}} = -\frac{1}{2\pi^2} g' g_{V}g_{\tilde{V}} &\Big[ \tilde{A}_1\, \epsilon_{\alpha \mu \nu \rho}\,p^\alpha+\tilde{A}_2\, \epsilon_{\alpha \mu \nu \rho}\,q^\alpha + {A}_3 \,\epsilon_{\alpha \beta \mu \rho} \, p^\alpha q^\beta p_{\nu}  + \\
&+  {A}_4\, \epsilon_{\alpha \beta \mu \rho} \, p^\alpha q^\beta q_{\nu} +{A}_5\, \epsilon_{\alpha \beta \nu \rho} \, p^\alpha q^\beta p_{\mu} +{A}_6\, \epsilon_{\alpha \beta \nu \rho} \, p^\alpha q^\beta q_{\mu} \Big] \;, \label{rosenberg}
\end{aligned}
\end{eqnarray}
where $p$ and $q$ are the EW gauge boson momenta, $g',g_{V},g_{\tilde{V}}$ are the gauge coupling constants, and the form factors $A_i$ with $i=3,\dots,6$ are fixed by the fermion loop diagrams and include the non-trivial dependence with external momenta,
\begin{equation}
    A_i = \sum_{f=\mu,\nu_\mu} t_f I_i(p^2,q^2,p\cdot q, m_f), \quad i=3,\dots,6\;,
\end{equation}
where $I_i$ are integrals over Feynman parameters of the loop, and we sum over the $U(1)'_{\mu}$-charged fermions in the EFT with their relevant combination of EW$\times U(1)'_{\mu}$ group charges $t_f$. The form factors $\tilde{A}_i$ in the loop amplitudes have been regularized by requiring the EW gauge symmetry to be anomaly free in the EFT~\footnote{The method of Rosenberg~\cite{Rosenberg:1962pp} exploits physical conditions in order to determine the regularized form factors $\tilde{A}_1$ and $\tilde{A}_2$ in term of the finite form factors $A_3,\dots,A_6$ which can be calculated directly in 4 dimensions. In this way one does not have to rely on dimensional regularization for the divergent form factors and bypasses the problem of defining $\gamma_5$ and  the antisymmetric Levi-Civita tensor in dimensions $d>4$ (for further details see, for example, the appendix B of~\cite{Dedes:2012me}).}, i.e.~by imposing the corresponding Ward identities~\cite{Racioppi:2009yxa},
\begin{align}
    p^\mu  A_{\rho \mu \nu}^{Z' V \tilde{V}} + i\, m_V A^{Z' G \tilde{V}}_{\rho\nu}&= 0\;, \label{eq:wardp}\\
    q^\nu  A_{\rho \mu \nu}^{Z' V \tilde{V}} + i\, m_{\tilde{V}} A^{Z' V \tilde{G}}_{\rho\mu}&= 0\;, \label{eq:wardq}
\end{align}
where the second terms on the left correspond to the Goldstone boson contributions, which are present only for massive gauge bosons and are calculated by replacing the external vector by its corresponding scalar Goldstone in the triangle diagram. This form is known as the covariant anomaly in the literature and corresponds to a specific choice in the WZW terms or in the shift-symmetry freedom of the loop momentum. The form factors $\tilde{A}_1$ and $\tilde{A}_2$ we obtain from Eqs.~(\ref{eq:wardp}) and (\ref{eq:wardq}) will generally depend on the nondivergent form factors $A_i$, external momenta contractions and, if present, Goldstone amplitudes. Therefore, the structure of the triple gauge couplings we want to simulate depends on the integrals over the Feynman parameters and has a complicated dependence on external momenta. The explicit form of the vertices we simulated are given in Appendix~\ref{appendix:couplings}.
The  $U(1)'_\mu$ mixed Ward identity reads,
\begin{equation}
    (p+q)^\rho A^{Z'V\tilde{V}}_{\rho\mu\nu} = -\frac{1}{2\pi^2} g' g_{V} g_{\tilde{V}} (\tilde{A}_1-\tilde{A}_2)\epsilon_{\alpha\beta\mu\nu}p^\alpha q^\beta ~,\label{eq:wardZp}
\end{equation}
with the coupling constants $g_V=e$ if $V=\gamma$ and $g_V=g_Z/2$ if $V=Z$,  where $g_Z = \sqrt{g_1^2+g_2^2}$, with $g_1$ and $g_2$ the SM coupling constants of $U(1)_Y$ and $SU(2)_L$, respectively, and for simplicity, in this expression, we are considering the light fermion mass $m_\mu=0$. 

Examining the vertex as expressed in Eq.~(\ref{rosenberg}) for the $Z'V\tilde{V}$ coupling and recalling that it is the longitudinal component of $Z'$ the one sensitive to the anomaly, as is clear via the anomalous Ward identity Eq.~(\ref{eq:wardZp}), one can see that only the transverse components of the SM gauge bosons in the vertex will provide a non-vanishing contribution at high energies. This can be checked explicitly in the signals of interest either when the $Z'$ is replaced by the longitudinal part of its propagator in $s$-channel production or when it appears as an on-shell final state particle via its longitudinal polarization.

In our analysis, we simulate both signal and background processes at parton-level with \texttt{MadGraph} \cite{Alwall:2014hca}, in which the model used to generate events is given in the Universal Feynrules Output (UFO) format \cite{Degrande:2011ua}. We first implement our model by adding the tree-level $Z'\mu\mu$ and $Z'\nu_\mu\nu_\mu$ couplings to the SM lagrangian via \texttt{Feynrules} \cite{Alloul:2013bka} and exporting it in UFO format. 
The anomalous couplings in our model include both the SM fermion loops and the non-decoupling effects, and therefore cannot be written in terms of a Lagrangian. 
Instead, we directly modify the UFO files to add the anomalous couplings with their corresponding form factors~\footnote{In the implementation of the form factors we neglect the Goldstone terms in the Ward identities (Eqs.~(\ref{eq:wardp}) and~(\ref{eq:wardq})) since those result to be proportional to $m_{\mu}^2$.}. 

In processes where the Lorentz-invariant factors $p^2$, $q^2$ and $p\cdot q$ are fixed ($p$ and $q$ stand for the EW gauge boson momenta), we can treat the loop integrals as external inputs, which can be computed by an external software and then feed them to \texttt{MadGraph} via the \texttt{param\_card} of the process. This is the case for on-shell diboson production at lepton colliders at a fixed energy. For more general processes we need to compute the $I_i$ integrals on the fly during event simulation for any given point in phase space, which is needed for hadron colliders, for example, in LHC simulations where partonic center-of-mass energy is not fixed and integration over the parton distribution functions is necessary. We construct custom functions for numeric evaluation of these integrals in \texttt{Fortran}, which are implemented into the UFO file structure by including the triple-boson vertices in the form of Eq.~\eqref{rosenberg} with the corresponding form factors. 
We also verified that the numerical results given by our code are consistent with more precise dedicated software such as \texttt{LoopTools} \cite{Hahn:1998yk}. The implementation of the latter, however, was much slower in conjunction with the event generation and therefore we decided to use our own code.

The underlying process and hadronization of partonic final states is simulated with \texttt{Pythia} \cite{Sjostrand:2014zea}. Fast detector simulation is performed with \texttt{Delphes} \cite{deFavereau:2013fsa} by using collider-specific cards. For LHC processes we use the default ATLAS card, for $e^+e^-$ collider we use the CLIC card included for the $3\;{\rm TeV}$ stage, and for the muon collider we use a hybrid of CLIC and FCC-hh cards that matches the current expected performance of the detectors \cite{mucoll}. In all of our analyses we simulate signal at LO in the anomalous couplings and consider only irreducible backgrounds. 
As we will see in the LHC case, this turns out to be an optimistic approach which however will not change the conclusions on their capabilities. 
For the futuristic hadronic and leptonic cases, it seems reasonable to consider only the irreducible backgrounds given the current uncertainties of the actual experimental setups.

As shown in~\cite{Medina:2021ram}, the strongest current bounds on the model under consideration come from neutrino trident production. Thus in order to maximize the cross-sections for the signal, we use throughout our analysis the largest $g_{\mu}$ value allowed by neutrino trident constraints \cite{CCFR:1991ccs, Altmannshofer:2019zhy}, given by
\begin{eqnarray}
g^{\rm max}_\mu \simeq \sqrt{0.3}~ \frac{m_{Z'}}{v}\;,
\label{eq:trident}
\end{eqnarray}
where $v \simeq 246 $ GeV is the electroweak symmetry breaking vacuum expectation value. For the values $m_{Z'} \in [100, 1000]$ GeV, we consider the couplings $g^{\rm max}_\mu \in [0.22, 2.2]$.

\section{Studies and implementation at hadron colliders}
\label{sec:hadroncoll}

It seems natural first attempting to probe the anomalous triple-gauge couplings at hadron colliders since the LHC is currently the only available high-energy collider taking data. Furthermore, it is easier to reach larger values for the center-of-mass energy in collisions at hadron colliders than at lepton colliders due to the small percentage energy lost to synchrotron radiation. In this section we will consider the High luminosity LHC (HL-LHC) at $\sqrt{s}=14$ TeV and integrated luminosity of $\mathcal{L}=3$~ab$^{-1}$ and the futuristic 100 TeV collider with a maximal integrated luminosity of $\mathcal{L}=20$~ab$^{-1}$, the latter as the highest energy hadron collider currently in consideration to be built in the future.

\subsection{Discovery prospects for the HL-LHC}
\label{sec:HL-LHC}

\begin{figure}[h]
    \centering
    \includegraphics[width=0.6\textwidth]{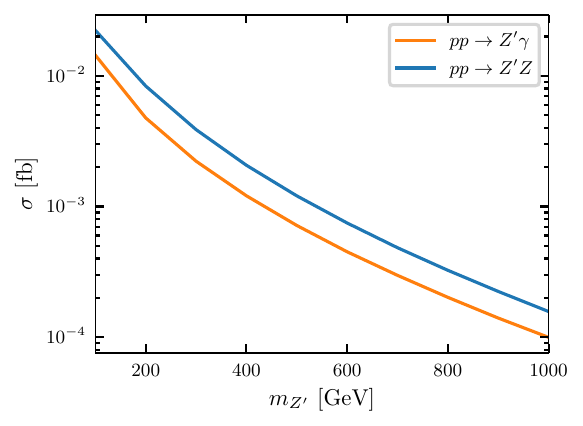}
    \caption{Production cross sections of $Z'$ and an EW boson at the LHC at $14~\mathrm{TeV}$. These processes are calculated from diagrams involving the anomalous gauge couplings. For each $m_{Z'}$ value $g_\mu$ is chosen to saturate the trident bound, see Eq.~\eqref{eq:trident}.}
    \label{fig:lhc_xsecs}
\end{figure}

At the LHC, anomalous couplings could potentially be probed in $pp\to \gamma^* / Z^* / W^* \to V Z'$ processes, where $V$ is an EW boson. 
We first use the UFO model described in Sect.~\ref{sec:model} to estimate the inclusive $VZ'$ production cross sections with \texttt{MadGraph}. Choosing $g_\mu$ to saturate the neutrino trident bound, Eq.~\eqref{eq:trident}, we show in Fig.~\ref{fig:lhc_xsecs} the cross section for $V=\gamma,Z$ in the $100$-$1000~\mathrm{GeV}$ range for the expected final center-of-mass collision energy of $\sqrt{s}=14\;{\rm TeV}$ at the LHC. 
In addition, our model also allows for $Z'W^\pm$ production, with the anomalous $Z'WW$ coupling being of similar size as the $Z'ZZ$ coupling. 
We expect a similar reach for both couplings and focus on the neutral channels. 
We observe that the $Z'Z$ cross section dominates across the explored mass range.
With the expected integrated luminosity value of $\mathcal{L}=3\;{\rm ab}^{-1}$ for the full LHC lifetime, we see that, in order to have at least $\mathcal{O}(10)$ signal events, $Z'$ has to be below $300~\mathrm{GeV}$.
In what follows, we define simple search strategies for each production channel in order to estimate the signal significance.

A standard search strategy for the anomalous couplings involves reconstructing both the $Z'$ boson through its decay products and the accompanying EW boson. 
The $Z'$ decay is dominated by the $\mu^+\mu^-$ channel, identifiable as a mass resonance, or it could decay into $\bar{\nu}_\mu \nu_\mu$ which produce missing transverse energy in the event. 
We focus in the dominant $Z'\to\mu^+\mu^-$ decay channel, since a selection window around the resonance invariant mass is expected to yield larger significances.
The search strategy also depends on the reconstruction of the EW boson, either by its direct observation if it is a photon, or through its decay products if it is a $Z$. In the following subsections, we show some estimated sensitivity prospects for the $ZZ'$ and $\gamma Z'$ channels at the HL-LHC. 
We work in an optimistic scenario with $m_{Z'}=200~{\rm GeV}$ and $g_\mu=0.445$ which saturates the neutrino trident bound.
In Sect.~\ref{sec:2muGammaLHC} we show prospects for the $Z'\gamma$ production in the $2\mu+\gamma$ decay channel, while in Sect.~\ref{sec:2mu2jLHC} we show our analysis for the $Z'Z$ production in the $2\mu+2j$ channel.

\subsubsection{$pp\to\gamma^*/Z^*\to Z' \gamma \to \mu^+\mu^-\gamma$ channel} 
\label{sec:2muGammaLHC}

With the setup described in Sect.~\ref{sec:model}, we simulate the signal $pp \to Z'\gamma \to \mu^+\mu^-\gamma$ and the irreducible backgrounds. 
In order to make the simulation more efficient, we impose $p_{T\gamma}>10~\mathrm{GeV}$, $p_{T\mu}>10~\mathrm{GeV}$, $|\eta_\gamma|<2.5$, $\Delta R_{\mu\mu}>0.4$, $\Delta R_{\gamma\mu}>0.4$ and $m_{\mu\mu}\in(100,300)~\mathrm{GeV}$ at parton level. 
After detector simulation, we apply the following cuts to both samples:
\begin{enumerate}
    \item Photon selection: at least $1$ photon with $p_{T\gamma}>60~{\rm GeV}$ and $|\eta_\gamma|<2.5$,
    \item Muon pair selection: at least $2$ muons with $p_{T\mu}>10~{\rm GeV}$, $|\eta_\mu|<2.4$ and $\Delta R_{\gamma\mu}>0.4$, forming pairs of opposite charges with $\Delta R_{\mu^+\mu^-}>0.4$, 
    \item $Z'$ reconstruction: at least one of the muon pairs has $|m_{\mu^+\mu^-}-m_{Z'}|<10~{\rm GeV}$.
\end{enumerate}
We estimate the expected number of signal ($S$) and background ($B$) events after applying these cuts and show the corresponding cutflow in Table~\ref{tab:cutflow-2muA}. We obtain an estimated significance of $S/\sqrt{B}\approx0.08$ for this analysis.

\begin{table}[]
\begin{center}
\begin{tabular}{l c c }
 \hline 
 \hline 
   & Signal & Background \\
 \hline
 Generator level cuts & $7.91$ & $1.17\times 10^{6}$ \\
 \hline
 Photon sel.  & $6.49$ & $6.27\times 10^{4}$ \\ 
 Muon pair sel.  & $3.83$ & $2.92\times 10^{4}$ \\ 
 $Z'$ reco.  & $3.20$ & $1.62\times 10^{3}$ \\ 
 \hline
 \hline
\end{tabular}
\end{center}
\caption{\label{tab:cutflow-2muA} Cutflow for the number of signal and background events in the $Z'\gamma$ channel at $14~{\rm TeV}$ at the LHC. Events are normalized using the calculated cross sections for $pp\to Z'\gamma\to\mu^+\mu^-\gamma$ including the parton level cuts, with a total integrated luminosity of $\mathcal{L}=3~{\rm ab}^{-1}$. Signal is simulated with $m_{Z'}=200~{\rm GeV}$ and $g_\mu=0.445$.}
\end{table}

In addition, interference between signal and background diagrams could potentially be sizable for this process since the $Z'$ considered is not too heavy and there could be interference in particular with the SM process $pp\to Z\gamma$ with a $t$-channel quark exchange. We compute these interference terms in \texttt{MadGraph} by generating $pp\to\mu^+\mu^-\gamma$ at a fixed order of $g_\mu^2$ in the squared amplitude of the process, which forces only interference terms in the cross section calculation. After applying the same cuts to the interference events as for the signal and background events, we obtain that interference terms are negative and of similar magnitude as the signal, which undermines even more our expectations for this channel at the LHC. 

\subsubsection{$pp\to\gamma^*/Z^*\to Z' Z \to \mu^+\mu^-jj$ channel}
\label{sec:2mu2jLHC}
We use again the setup described in Sect.~\ref{sec:model} and simulate the signal $pp \to Z'Z \to \mu^+\mu^-jj$ along with the irreducible backgrounds.
In order to make the simulation more efficient, we apply the following parton level cuts: $p_{Tj}>20~\mathrm{GeV}$, $p_{T\mu}>10~\mathrm{GeV}$, $|\eta_j|<5$, $|\eta_\mu|<2.5$, $\Delta R_{\mu\mu}>0.4$, $\Delta R_{j\mu}>0.4$, $\Delta R_{jj}>0.4$, $m_{jj}\in(70,110)~\mathrm{GeV}$  and $m_{\mu\mu}\in(100,300)~\mathrm{GeV}$, where $j$ stands for a quark or gluon in the partonic final state. During detector simulation, jets are reconstructed with the anti-$k_T$ algorithm with $\Delta R=0.5$. We impose the following requirements to both samples:

\begin{enumerate}
    \item Jets selection: at least $2$ reconstructed jets with $p_{Tj}>20~{\rm GeV}$, $|\eta_j|<5$ and $\Delta R_{jj}>0.4$,
    \item Muon pair selection: at least $2$ muons with $p_{T\mu}>10~{\rm GeV}$, $|\eta_\mu|<2.4$ and $\Delta R_{j\mu}>0.4$, forming pairs of opposite charges with $\Delta R_{\mu^+\mu^-}>0.4$,
    \item $Z$ reconstruction: at least one pair of jets with $|m_{jj}-m_{Z}|<10~{\rm GeV}$,
    \item $Z'$ reconstruction: at least one muon pair with $|m_{\mu^+\mu^-}-m_{Z'}|<10~{\rm GeV}$.
\end{enumerate}

Note that the third cut is necessary to reduce the large QCD background from dijet production, however also suppressing possible vector boson fusion contributions via triple gauge anomalous couplings.

A cutflow in terms of the number of events is given in Table~\ref{tab:cutflow-2mu2j}. The estimated significance for this analysis is $S/\sqrt{B}\approx0.01$. We also investigate if interference effects are sizable in the $Z'Z$ channel  and find that in this case the interference is also destructive and of the order of 10$\%$ of the signal after applying the cuts.
\begin{table}[]
\begin{center}
\begin{tabular}{l c c }
 \hline 
 \hline 
   & Signal & Background  \\ 
 \hline
 \hline
 Generator level cuts  & $8.33$ & $2.98\times 10^{6}$  \\ 
 \hline
 Jets sel.  & $6.88$ & $2.50\times 10^{6}$  \\ 
 Muon pair sel.  & $3.71$ & $1.52\times 10^{6}$ \\ 
 $Z$ reco.  & $1.63$ & $6.02\times 10^{5}$ \\
 $Z'$ reco.  & $1.43$ & $1.28\times 10^{4}$ \\ 
 \hline
 \hline
\end{tabular}
\end{center}
\caption{\label{tab:cutflow-2mu2j} Cutflow for the number of signal and background events in the $pp\to Z'Z\to\mu^+\mu^-jj$ channel at $14~{\rm TeV}$ at the LHC. Events are normalized using the estimated cross sections with a total integrated luminosity of $\mathcal{L}=3~{\rm ab}^{-1}$. Signal is simulated with $m_{Z'}=200~{\rm GeV}$ and $g_\mu=0.445$.
}
\end{table}

It is clear from the results in the different channels investigated that the HL-LHC is incapable of probing the anomalous triple-gauge couplings. 
Even when the signal-to-background ratio is improved by more than two orders of magnitude after applying the search strategy (for the $Z'\gamma$ channel), the extremely large initial difference between the small number of signal events (due to the loop nature of the involved couplings) and the large backgrounds leads to negligible significances.
Lastly, we analyze the possibilities of what is at the moment the highest energy hadron collider under consideration, the $100~{\rm TeV}$ proton-proton  collider.

\subsection{The 100 TeV proton-proton collider capabilities}
\label{sec:100TeV}

There have been many studies involving the research and development (R$\&$D) of a  $pp$ collider at a collision energy of $\sqrt{s}=100$ TeV as well as potential BSM signals that this collider may probe~\cite{Arkani-Hamed:2015vfh, Mangano:2016jyj}. In what follows we consider the potential of the 100 TeV collider in probing anomalous triple-gauge couplings, using the \texttt{Delphes} FCC-hh card~\cite{Selvaggi:2717698}.

\begin{figure}
    \centering
    \includegraphics[width=0.7\linewidth]{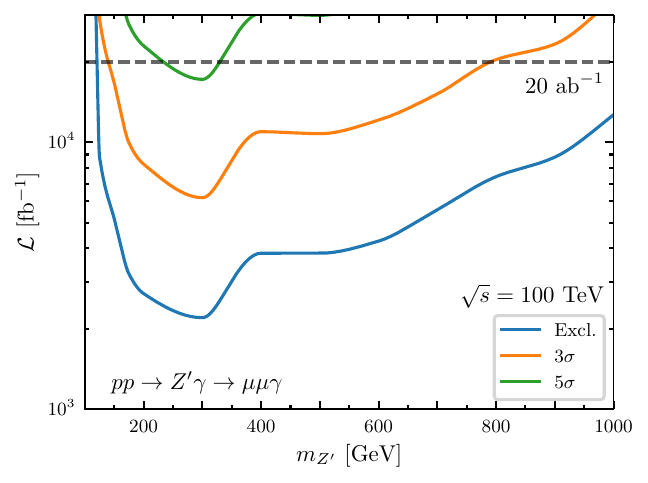}
    \caption{Integrated luminosity required for exclusion, $3\sigma$ and $5\sigma$ significance for $pp\to Z'\gamma \to \mu^+\mu^-\gamma$ with $\sqrt{s}=100~\mathrm{TeV}$. The dashed black line represents the estimated maximal luminosity of $20~\mathrm{ab}^{-1}$ for the FCC-hh.}
    \label{fig:lumi100TeV}
\end{figure}

Applying the same search strategy as in the LHC for the $Z'Z$ channel, we obtain a significance of $S/\sqrt{B}\approx0.17$ for $m_{Z'}=200~\mathrm{GeV}$ at $\mathcal{L}=20~\mathrm{ab}^{-1}$ and similar small significances for other $Z'$ masses. 
In contrast, for the $Z'\gamma$ channel, a substantial improvement can be made due to the larger available energy, by demanding a higher $p_T$ cut on the photon than in the previous LHC study for this channel.
We scan over $m_{Z'}$ between $100~\mathrm{GeV}$ and $1000~\mathrm{GeV}$ with $g_\mu$ satisfying the neutrino trident bound, applying the same cuts as in the LHC analysis except for the cut in the $p_T$ of the photon, which is chosen to maximize the signal significance, and the di-muon invariant mass window. 
For the latter cut, we choose $|m_{\mu^+\mu^-}-m_{Z'}|<10~\mathrm{GeV}$ for $m_{Z'}\leq 400~\mathrm{GeV}$, and $|m_{\mu^+\mu^-}-m_{Z'}|<0.1 \,m_{Z'}$ for $m_{Z'}>400~\mathrm{GeV}$.
This is consistent with the narrow resonance at lower masses, while enhancing the signal acceptance for heavier masses that exhibit a broader resonance. 
In addition, at generator level we simulate events with a cut in $p_{T\gamma}$ in order to make the simulation process more efficient.
We ensure that the final $p_{T\gamma}$ cut adopted in the analysis is at least $100~\mathrm{GeV}$ larger than the generator level cut.
The background is strongly suppressed by the cuts of the search strategy which prevents the computation of the corresponding efficiency with enough precision unless a very large amount of events is simulated. Instead, we adopt a conservative approach for the estimation of the background efficiencies, by adding $+1\sigma$ to the number of simulated events that pass the detector level cuts $B_{\rm sim}$, where $\sigma=\sqrt{B_{\rm sim}}$.
The values of $p_{T\gamma}$ cuts that maximize the significance are between $p_{T\gamma}>300~\mathrm{GeV}$ for $m_{Z'}=100~\mathrm{GeV}$ and $p_{T\gamma}>1700~\mathrm{GeV}$ for $m_{Z'}=1000~\mathrm{GeV}$.
Luminosities required for exclusion, evidence, and discovery level significances are shown in Fig.~\ref{fig:lumi100TeV}.
We see that the lowest necessary luminosities for any of the significance levels are achieved for $m_{Z'}\approx300~\mathrm{GeV}$. 
Note that in the range of $m_{Z'}\in[230,330]$ discovery significances would be possible at the maximum projected luminosities of the 100 TeV $pp$ collider. 
At the same luminosity, evidence can be found for $m_{Z'}\in[150,800]$ GeV and we could exclude almost the whole range of $Z'$ masses analyzed. For $m_{Z'}\in[100,200]$ GeV the background starts to increase, in particular as we move closer to the $Z$ gauge boson mass due to the $Z\gamma$ background, implying a larger necessary luminosity. 
For masses near $m_{Z'}\approx400~\mathrm{GeV}$, both prescriptions for the di-muon mass window are inefficient, either too sharp or too wide when compared with the total width of $\Gamma_{Z'}\approx12~\mathrm{GeV}$, leading to an increase in the required luminosity and a step in the curves.
A more adequate choice of window width could partially reduce the required luminosity and soften the transition between the two regimes, however this would correspond to an actual fine tuning and we decide not to pursue it further. 
For larger masses, we observe that by taking an invariant mass window centered at the $Z'$ resonance but whose size increases with $m_{Z'}$,  we are able to mitigate an otherwise sharp increment in the required luminosity due to the smaller values of the signal cross section at larger masses.

A cutflow for $m_{Z'}=300~\mathrm{GeV}$ is shown in Table~\ref{tab:100TeVcoll}.
Generator level cuts include $p_{T\gamma}>400~\mathrm{GeV}$ in both signal and background events, and photon selection cuts include $p_{T\gamma}>650~\mathrm{GeV}$.
The significance for discovery in this case is $5.4$ for the maximal expected luminosity of $\mathcal{L}=20~\mathrm{ab}^{-1}$ and evidence could be obtained for $\mathcal{L}=6~\mathrm{ab}^{-1}$ .

\begin{table}[h]
\begin{center}
\begin{tabular}{l c c }
 \hline 
 \hline 
   & Signal & Background \\ 
 \hline
 Generator level cuts & $156$ & $6.51\times10^4$ \\ 
 \hline
 Photon sel.  & $78.2$ & $7.80\times10^3$ \\
 Muon pair sel.  & $66.6$ & $6.28\times10^3$ \\
 $Z'$ reco.  & $51.7$ & $76.4$ \\
 \hline
 \hline
\end{tabular}
\end{center}
\caption{\label{tab:100TeVcoll} Cutflow for $pp\to Z' \gamma \to \mu^+\mu^-\gamma $ at  $\sqrt{s}=100~{\rm TeV}$ and total integrated luminosity of $\mathcal{L}=20~{\rm ab}^{-1}$. Signal is simulated at $m_{Z'}=300~\mathrm{GeV}$ and $g_\mu$ saturating the neutrino trident bound.}
\end{table}

\section{Studies and implementation at lepton colliders}\label{sec:leptoncoll}

Lepton colliders provide a cleaner environment in comparison with hadron colliders due to the weakness of the EW interactions and moreover, since leptons are elementary particles~\footnote{Due to the electromagnetic cloud surrounding charged leptons, they can also be studied using the PDF formalism, see \cite{Han:2020uid,Buonocore:2020nai,DaRold:2024ram} as a phenomenological example in colliders. We checked that these effects are negligible in our case.}, there is a much better handle in the energy of the colliding particles which is basically fixed by construction. As we will see, the latter capability will be crucial in probing the anomalous triple gauge couplings. We consider two types of lepton colliders: muon ($\mu^{+}\mu^{-}$) and electron-positron ($e^{+}e^{-}$) colliders. Both are futuristic in the sense that the colliding energies and luminosities considered have not been accomplished for $e^{+}e^{-}$ colliders and no muon collider has ever been built due to the technical difficulty in what is known as the cooling of muons. There have been however serious considerations and studies pushing forward the R$\&$D for both types of lepton colliders and their construction after the end of the LHC era is quite feasible~\cite{Accettura:2023ked,Black:2022cth,Delahaye:2019omf,Bartosik:2020xwr,MuonCollider:2022nsa,CLIC:2018fvx,Aicheler:2018arh}. 
In what follows, we show that both colliders can be useful in probing the anomalous triple gauge coupling for the model under consideration\footnote{We verified that interference between signal and SM is negligible, and therefore we do not take it into account in the following.}, in different $m_{Z^\prime}$ mass ranges, implying an advantage over the hadron collider counterpart. 

\subsection{ Muon colliders expectations at probing anomalous triple-gauge couplings}
\label{sec:muoncoll}

A current candidate for a sub-TeV muon collider is at the Higgs mass, $\sqrt{s}=m_H\simeq 125~{\rm GeV}$, where resonant Higgs production  is dominant and several of its decay channels can be studied with precision \cite{deBlas:2022aow, MuonCollider:2022nsa}. We showed in our previous work that in such facility it is also quite easy (with a $\mathcal{L}\sim \mathcal{O}(10)$ fb$^{-1}$) to discover a $Z'$ coupled to muons via tree-level couplings in the $\mu^+\mu^-\to Z^{\prime*}\to\mu^+\mu^-$ channel \cite{Medina:2021ram}.
Furthermore, it would be in principle also possible to test the non-vectorial nature of the $Z'\mu\mu$ couplings by measuring the forward-backward asymmetry of the muon pair produced. If such asymmetry is discovered, it would strengthen the hypothesis of a non-zero axial coupling and the possibility of anomalous triple-gauge couplings. 

In order to assess the exclusion and/or discovery reach of the anomalous gauge couplings at a muon collider, we explore different signals given by $ZZ$ or $Z\gamma$ production and their possible decay channels. After generating signal and background events with the setup described in Sect.~\ref{sec:model}, we apply some basic cuts depending on the target final state, as described below.

\begin{figure}
    \centering
    \includegraphics[width=0.7\linewidth]{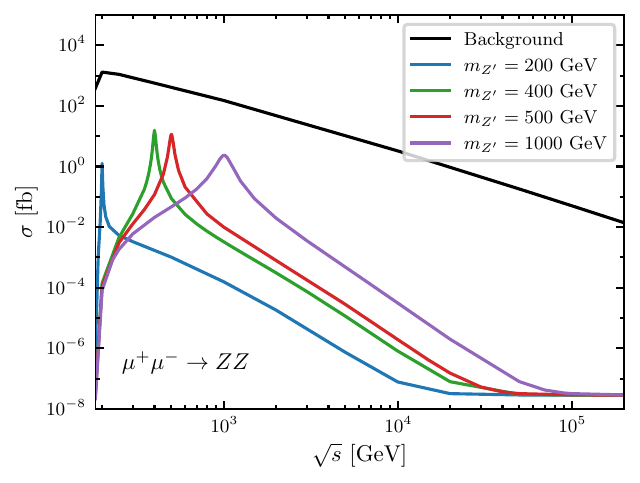}
    \caption{Cross section of $\mu^{+}\mu^{-}\to ZZ$ as a function of $\sqrt{s}$ for different $Z'$ masses, and $g_\mu$ set to saturate the neutrino trident bound for each mass. Background coming from SM is simulated at LO. }
    \label{fig:xsec_ZZ}
\end{figure}

We simulate $ZZ$ production events at parton level, via an $s$-channel $Z^{\prime}$, for different $Z'$ masses and colliding muon energies ($\sqrt{s}$) in order to estimate the inclusive cross section of this process, as well as the irreducible SM background at LO, $\mu^{+}\mu^{-}\to ZZ$. 
Results are shown in Fig.~\ref{fig:xsec_ZZ}, where the $ZZ$ production cross section is calculated with \texttt{MadGraph} as a function of $\sqrt{s}$ for four different $m_{Z'}$ values, and fixing $g_{\mu}$ to the maximum value allowed by the neutrino trident bound, see Eq.~\eqref{eq:trident}. 
In this plot one can see the resonance peak at $\sqrt{s}=m_{Z'}$, then for intermediate energies there is a power-law decay up to roughly $\sqrt{s}=100 \,m_{Z^{\prime}}$, where the cross sections reach a constant value that depends on the $Z'$ mass and the muon mass as $\sigma_{ZZ}\propto m^{2}_{\mu}/m^4_{Z'}$. In fact, due to the choice of the coupling $g_{\mu}$ saturating the trident bound, all benchmarks go to the same constant cross section asymptotic value. This behavior can be easily understood by calculating, respectively, the longitudinal ($\sigma_L$), transverse ($\sigma_T$) and interference contributions ($\sigma_{LT}$) from the $Z'$ propagator to the $ZZ$ production, which are given by

\begin{eqnarray}
\sigma_L = \frac{g_{\mu}^4 g_Z^4}{2 (4 \pi)^5} \left(\tilde{A}_1 - \tilde{A}_2 \right)^2 \frac{m_{\mu}^2}{m_{Z'}^4} \frac{\left( 1-4 m_Z^2/s \right)^{3/2}}{\sqrt{1-4m^2_{\mu}/s} \left(1-m^2_{Z'}/s \right)^2},\label{Eq.crossL}
\end{eqnarray}
\begin{eqnarray}
\sigma_{LT} = -\frac{g_{\mu}^4 g_Z^4}{2 (4 \pi)^5} \left(\tilde{A}_1 - \tilde{A}_2 \right)^2 \frac{m_{\mu}^2}{m_{Z'}^2 s} \frac{\left(1-4 m_Z^2/s \right)^{3/2}}{\sqrt{1-4m^2_{\mu}/s}\left(1-m^2_{Z'}/s \right)^2},\label{Eq.crossLT}
\end{eqnarray}
\begin{eqnarray}
\begin{aligned}
\sigma_T &= \frac{g_{\mu}^4 g_Z^4}{6 (4 \pi)^5} \frac{\left(1-4 m_Z^2/s \right)^{3/2} }{\sqrt{1-4m^2_{\mu}/s} \left(1-m^2_{Z'}/s \right)^2} \left[ 3 A_3^2 m_{\mu}^2 + \frac{4 m_{\mu}^2 m_Z^2 (A_3 + A_5)^2 (6m_Z^2 +m_{Z'}^2)}{s^2} \right. \\
 &+ m_Z^2 (A_3 + A_5)^2 - \left. (A_3 + A_5) \frac{m_Z^2}{s}((m_{Z'}^2+3m_Z^2)(A_3+A_5)+4(4 A_3+A_5) m_{\mu}^2)  \right].\label{Eq.crossT} \\
 \end{aligned}
\end{eqnarray}

\begin{figure}
\center
\includegraphics[width=0.7\linewidth]{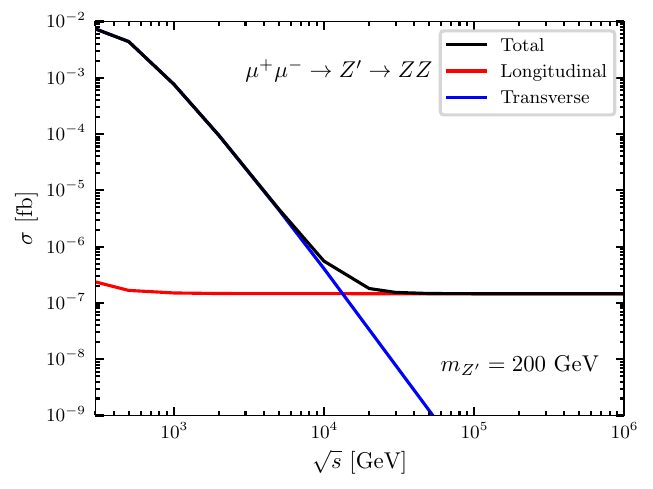}
\caption{ \label{fig:xsec_mumu200}{Cross section for the process $\mu^{+}\mu^{-} \to Z'^* \to Z Z$ with $m_{Z'}=200$ GeV and $g_{\mu} = 0.445$. We 
plot in blue (red) the transverse (longitudinal) contributions from the $Z'$ propagator, and in black the total cross section.
}}
\end{figure}

Analyzing these expressions in the $\sqrt{s}\gg m_{Z'}$ limit and using that in such limiting case $\tilde{A}_1$ and $\tilde{A}_2$ scale as constants while $A_3$ and $A_5$ scale as $1/s$ (see Appendix~\ref{appendix:couplings}), one can see that the transverse and interference contributions scale as $\sigma_T\sim m^2_{Z}/s^2$ and $\sigma_{LT}\sim m^2_{\mu}/(m^2_{Z'}\times s)$, while the longitudinal contribution goes to a constant $\sigma_L\sim m^{2}_{\mu}/m^4_{Z'}$. 
We see that the interference term is suppressed by the muon mass and $s$ and thus can be neglected. 
The transverse term while it is suppressed by $s^2$, it is proportional to $m^2_{Z}$, whereas the longitudinal term goes quickly to a small constant since it is proportional to $m^{2}_{\mu}$ \footnote{For the appearance of $m^{2}_{\mu}$ is crucial the axial nature of the coupling of the $Z'$ gauge boson to the muons.}. So what ends up happening is that after the resonant peak, the transverse component dominates but quickly drops as $1/s^2$ until its value is similar to the constant longitudinal contribution which then dominates the total cross section. This behavior is illustrated in Fig.~\ref{fig:xsec_mumu200} for $m_{Z'}=200$ GeV. Similar conclusions are obtained for the $Z\gamma$ final state.

The cross section behavior cannot be extended up to arbitrary high energies since, as described in Sect.~\ref{sec:model}, our EFT is equipped with a cut-off related to the energy at which the spectator fermions kick in. In anomaly-free theories the cross section is generically expected to decrease with $s$ to some power and since the SM is an anomaly-free theory, a similar dependence at large energies is expected for SM backgrounds. 
Though the case of a constant signal cross section along with background cross sections that drop with powers of $s$ is encouraging for probing our model at large energies, it turns out that the anomalous signal cross section tends to stabilize at energies that are too large and correspond to values too small to be probed at a muon collider. 

There is an interesting alternative in the case that the $Z'$ shows up in the final states $Z'Z$ or $Z'\gamma$, produced through an $s$-channel mediated by $Z$ or $\gamma$. Focusing in the $Z' Z$ final state, and considering the longitudinal polarization for the $Z'$ which should dominate at large $s$ due to the anomaly, we obtain the following expressions for the $Z'_L Z$ production cross section mediated by $\gamma$ and $Z$
\begin{eqnarray}
\sigma_{\gamma^{*},\;L} = e^4 g_{\mu}^2 g_Z^2 \left(\tilde{A}_1 - \tilde{A}_2 \right)^2 \frac{  2 \left( (1-m_Z^2/s+m_{Z'}^2/s)^2 -4 m_{Z'}^2/s \right)^{3/2} }{3 (4 \pi)^5 m_{Z'}^2},\label{fotonZL}
\end{eqnarray}
\begin{eqnarray}
\sigma_{Z^{*},\; L} = g_{\mu}^2 g_Z^6 \left(\tilde{A}_1 - \tilde{A}_2 \right)^2 \frac{(1-4 S_W^2 + 8 S_W^4) \left( (1-m_Z^2/s+m_{Z'}^2/s)^2 -4 m_{Z'}^2/s \right)^{3/2} }{48 (4 \pi)^5 m_{Z'}^2 \left(1-m^2_{Z}/s \right)^2 },\label{ZetaZL}
\end{eqnarray}
where $S_W = \sin{\theta_W}$, with $\theta_W$ the weak mixing angle. Notice that both contributions provide in the $\sqrt{s}\gg m_{Z'}$ limit a constant cross section that scales as $\sigma_L\propto 1/m^2_{Z'}$, with no muon mass suppression as was the case for an intermediate $Z'$. The limiting cross sections is indeed larger and it is reached at smaller $\sqrt{s}$ values than the previous asymptotic value obtained when the $Z'$ was the intermediate state, opening the possibility of detection at reasonable values of $\sqrt{s}$ for which the SM background have already dropped enough. Similar conclusions can be drawn for the $Z'\gamma$ final state. Unfortunately, in addition to the production of $Z'Z$ and $Z'\gamma$ mediated by the anomalous triple-gauge bosons couplings, there is also a new physics (NP) tree-level contribution with a muon exchanged in the $t$-channel that completely dominates the cross section. 
Since in our model the $Z^\prime$ only couples to muons (and muonic neutrinos), there is no such tree-level contribution in a $e^{-}e^{+}$ collider, which renders it a unique tool to probe the anomalous triple-gauge bosons couplings as we will see later.
In the next subsection we show that muon colliders still can be a powerful tool if the $Z'$ is resonantly produced. 
We work under the assumption that an anomalous $Z'$ coupled to muons has been discovered, either at the LHC or at a sub-TeV muon collider, and we consider a muon collider in which the collision energy is tuned to the $Z'$ mass. 
Besides providing the maximum value for the signal cross section, the $Z'$ production at resonance and its subsequent decay via the anomalous triple-gauge coupling is independent of $g_{\mu}$~\footnote{The cross section depends only on BR$(Z'\to \mu^{+}\mu^{-})$ and BR$(Z'\to CD)$ with $CD$ either $ZZ,\; Z\gamma$, both branching ratios independent of $g_{\mu}$~\cite{Medina:2021ram}.}, so that constraints on this coupling have no impact on the prospects of the resonant search.

\subsection{Resonant production at muon colliders}

In the following we explore the $ZZ$ production in a muon collider at the $Z'$-resonance, $\sqrt{s}=m_{Z'}$. Although we will show plots with larger luminosities, we will take as a sensible choice for the maximum luminosity attainable at a muon collider the value of 1 ab$^{-1}$ ~\cite{Delahaye:2019omf}, which should roughly correspond to a muon collider running for 20 years at energies of order the $Z'$ masses considered in this work. 
Among the possible decay channels of the $Z$ bosons, we concentrate on $Z\to jj$ and $Z \to e^+ e^-$. 
Decays to $\tau^+ \tau^-$ have a branching ratio comparable to $e^+ e^-$ but may suffer from lower tau reconstruction efficiencies with respect to electrons, and therefore are expected to yield less significant results.
Furthermore, final states that involve $\mu^+\mu^-$ are suppressed with respect to tree-level diagrams that produce one or two $Z'$ decaying into muons.
Since we want to study signals in which the main NP contribution arises from anomalous decays, we ignore this decay channel. 
Finally, invisible $Z\to \nu\bar{\nu}$ decays are also possible, resulting in missing energy in the process, but since $Z$ reconstruction is needed in order to characterize the anomalous couplings we do not consider these decay channels. We are left with three possible final states: $4j$, $4e$ and $2e2j$. We analyze each channel separately, simulating the signal and irreducible backgrounds as described in Sect.~\ref{sec:model}.  

\begin{figure}
    \centering
    \includegraphics[width=0.7\linewidth]{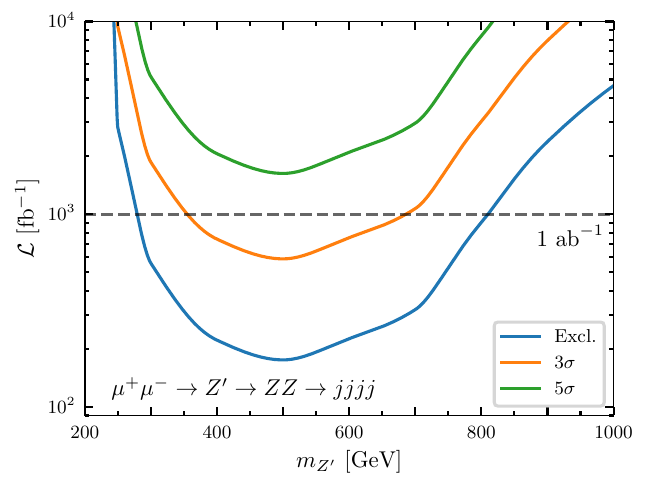}
\caption{Luminosity required for exclusion, $3\sigma$ and $5\sigma$ significance for resonant $\mu^+\mu^-\to Z' \to  ZZ\to\, 4j$ production. }
    \label{fig:lumi_ZZ_4j}
\end{figure}

For the $4j$ final state we scan over $\sqrt{s}=m_{Z'}$ between $200~\mathrm{GeV}$ and $1000~\mathrm{GeV}$. Jets are reconstructed using the $k_T$ algorithm with $R=0.5$, and we apply the following cuts:
\begin{enumerate}
    \item At least $4$ jets with $p_{Tj}>20~{\rm GeV}$, $\eta_j < 5$,
    \item At least $2$ pairs of jets satisfying $|m_{jj}-m_Z|<10~{\rm GeV}$.
\end{enumerate}

Luminosities required for exclusion, evidence and discovery level significances are shown in Fig.~\ref{fig:lumi_ZZ_4j} in terms of $m_{Z'}$. We see that for the three significance levels considered the minimum required luminosity is reached at $m_{Z'}\approx 500$ GeV, and the sensitivity of this search quickly degrades for small masses due to the increase in the background cross section and the decrease in the signal cross section close to the $ZZ$ production threshold. For large masses the sensitivity also drops, but more slowly since in this case the background cross section is decreasing. With a maximum luminosity of 1 ab$^{-1}$ it is possible to exclude masses between 280 GeV and 800 GeV, and reach evidence level in the $m_{Z'}$ range of roughly 380 GeV  to 700 GeV. Finally, discovery level significance seems  unreachable at the maximum estimated luminosity. 

We provide an example of the cutflow for $m_{Z'}=500$ GeV in Table~\ref{tab:my_label} for a $\mathcal{L}=1$ ab$^{-1}$. Although the signal cross section is larger at $m_{Z'}=400$ GeV, the smaller background and in particular the acceptance after cuts imply a larger significance for the signal at $m_{Z'}=500$ GeV.  
\begin{table}[t]
    \centering
    \begin{tabular}{c c c c c}
    \hline \hline
        & Signal & Background & Rel. acc. sgnl & Rel. acc. bkg \\  
         \hline
         Initial & $4.95\times 10^3$ & $2.28\times 10^6$ & - & - \\
        \hline
         $4j$ sel. & $3.02\times 10^3$ & $1.25\times 10^6$ & $0.610$ & $0.547$ \\
         $Z$ windows & $1.02 \times 10^3$ & $6.70\times 10^4$ & $0.336$ & $0.0537$ \\
 \hline \hline
    \end{tabular}
    \caption{Cutflow for $\mu^+\mu^-\to ZZ\to4j$ searches at $m_{Z'}=\sqrt{s}=500~\mathrm{GeV}$ and luminosity $\mathcal{L}=1\;{\rm ab}^{-1}$. Cuts are described in the main text. Initial number of events and number of events surviving each cut are provided for signal and background in second and third columns. Relative acceptances for signal and background are given in fourth and fifth columns respectively.}
    \label{tab:my_label}
\end{table}

Another possible decay product of the pair $Z Z$ are a pair of electrons and a pair of jets, the $2e2j$ channel. This channel has the advantage that it is relativity easy to reconstruct: an electron pair is expected to come from a $Z$ decay, and the jets from the other one. 
The SM backgrounds for the process $ \mu^+ \mu^- \to Z Z$ are important, but the main contribution comes from a $t$-channel exchange of a muon, whereas our signal is $s$-channel. This suggests the possibility of use a cut on the pseudo-rapidity $\eta$ of the electron pair, to exploit the different angular distribution of the signal.

\begin{figure}[h]
\center
\includegraphics[width=0.7\linewidth]{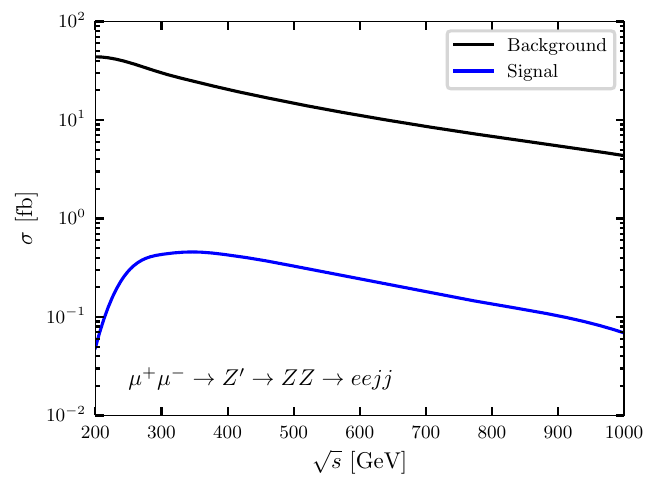}
\caption{ \label{fig:xsec_ejj}{Signal and background cross sections for resonant $\mu^+\mu^-\to Z' \to ZZ\to e^+ e^- jj$ production.} }
\end{figure}

The signal cross sections for the resonant production of the $Z'$, decaying to a pair $ZZ$ with semi-leptonic decay are shown in Fig.~\ref{fig:xsec_ejj}, with the corresponding SM background. We can see that the signal cross section is almost two orders of magnitude lower than the background cross section and is maximal for the masses $m_{Z'} = 300 \sim 400$ GeV.

For the analysis we applied the following cuts:
\begin{itemize}
\item Selection cuts: at least 1 jet with $p_{Tj} > 20 $ GeV and $|\eta_j| < 5$, at least 1 $e^+$ and 1 $e^-$ with $p_{T \ell} > 10 $ GeV and $|\eta_{\ell}| < 5$.

\item At least one pair of electrons with $|m_{e^+e^-}-m_Z|<10 $ GeV and $|\eta_{e^+e^-}|<1$.

\item Invariant mass of the sum of jets (hadronic) $30$ GeV $<m_{jets}< 110 $ GeV.

\end{itemize}

As an illustration, we provide an example of the cutflows for the mass of $m_{Z'} = 400$ GeV in the Table \ref{tab:tabla_400_eej} for the dedicated search of the $ e e jj$ signal in the Muon Collider. The total integrated luminosity is set at $1$ ab$^{-1}$. 

\begin{table}
\begin{center}
\begin{tabular}{l c c }
\hline 
\hline 
 & Signal & Background \\ 
\hline 
Initial                           & 427.3  & 19165.0 \\ 
\hline 
Selection cuts                    & 294.0  & 11608.2 \\ 
$|m_{e^+e^-}-m_{Z}| < 10 $ GeV       & 259.2  & 7991.8 \\ 
$|\eta_{e^+e^-}| < 1$                & 191.4  & 3376.9 \\ 
$30 $ GeV $< m_{jets} < 110 $ GeV & 123.3  & 1929.9 \\ 
\hline 
\hline 
\end{tabular}
 \end{center}
\caption{\label{tab:tabla_400_eej}  
Cutflow for the number of signal and background events for the resonant channel $\mu^+ \mu^- \to Z' \to Z Z \to e^+ e^- j j$ at the muon collider. Events are normalized using the estimated cross sections with a total integrated luminosity of $\mathcal{L}=1~{\rm ab}^{-1}$. Signal is simulated with $m_{Z'}=400~{\rm GeV}$. The selection cuts are described in the text.}
\end{table}

\begin{figure}[h]
\center
\includegraphics[width=0.7\linewidth]{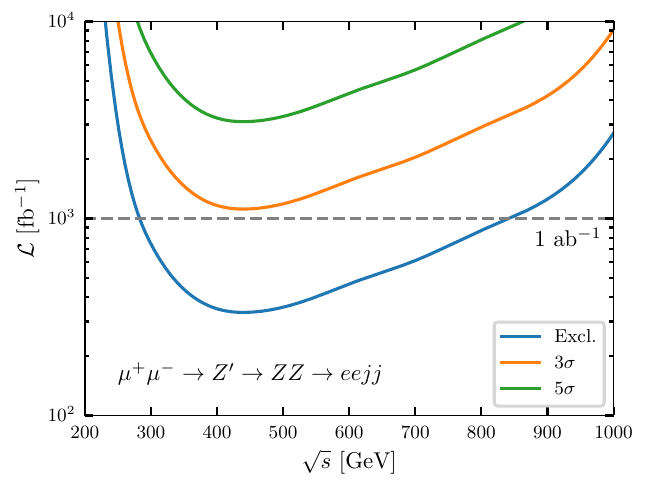}
\caption{ \label{fig:z_ejj_lumi}{Luminosity required for exclusion, $3 \sigma$ and $5 \sigma$ significance for resonant $\mu^+\mu^-\to Z' \to ZZ\to e^+ e^- jj $ production.} }
\end{figure}

We show in Fig.~\ref{fig:z_ejj_lumi} for the resonant $ZZ$ production decaying into $e^+ e^-$ plus two jets, as function of $\sqrt{s}=m_{Z'} $, the required luminosities for exclusion, $3\sigma$ evidence and $5\sigma$ discovery. In this channel, note that a luminosity of $3$ ab$^{-1}$ would be required to achieve a $3 \sigma$ significance for $Z'$ masses ranging from $300$ GeV to $800$ GeV and a luminosity of $4$ ab$^{-1}$ would be needed to reach the coveted $5 \sigma$ discovery threshold for $Z'$ masses between $400$ GeV and $600$ GeV. These values for the luminosities are most likely beyond the capability reach of a muon collider running at $\sqrt{s}=m_{Z'}$ , for the $Z'$ masses considered. For the more sensible choice of maximal luminosity of 1 ab$^{-1}$ we see that we could in principle only put exclusion limits for $m_{Z'}\in [280, 850]$ GeV. 

Finally, for the $4e$ final state we perform an analysis similar to the $4j$ case, simulating signal and background and selecting events that contain $2$ pairs of opposite-sign electrons that satisfy $|m_{e^+e^-}-m_Z|<10~{\rm GeV}$. 
We see that, due to the low branching ratio of $Z\to ee$ compared to $Z\to jj$, the significance is not higher than $S/\sqrt{B}\approx 0.36$ for $m_{Z'}=400\;\mathrm{GeV}$.

We do not consider the channels with resonant production of $Z \gamma$ because its cross sections are about four orders of magnitude lower than their respective backgrounds, and possible cuts are not efficient enough to get a significance larger than $0.8\sigma$ with a total luminosity of $1$ ab$^{-1}$, in particular due to the irreducible SM $Z\gamma$ background. As we will see in the next subsection, this changes if instead of the $Z$ boson in the final state we have the $Z'$, making the invariant mass cut window around $m_{Z'}$ more efficient in discriminating against the background.

\subsection{Non-resonant production at $e^+ e^-$ collider}
\label{sec:eecoll}

As mentioned previously, anomalous triple-gauge couplings can also be probed at future electron-positron colliders, such as the proposed FCC-ee \cite{FCC:2018evy,dEnterria:2016sca}, ILC \cite{Behnke:2013xla, Baer:2013cma}, CLIC \cite{CLIC:2018fvx, Aicheler:2018arh} and CEPC \cite{CEPCStudyGroup:2018rmc,CEPCStudyGroup:2018ghi}.
In these colliders, the processes $e^+ e^- \to Z^*/\gamma^* \to Z' Z$ and $e^+ e^- \to Z^*/\gamma^* \to Z' \gamma$, with the $Z'$ decaying to a pair of muons, provide potential windows to explore triple gauge couplings.
The $Z'$ production is via  non-resonant process, and thus the signal cross-section depends on the coupling $g_{\mu}$.
As mentioned before we use the largest $g_{\mu}$ value allowed by neutrino trident constraints, as described in Eq.~(\ref{eq:trident}).

\begin{figure}[t]
\center
\includegraphics[width=0.495\linewidth]{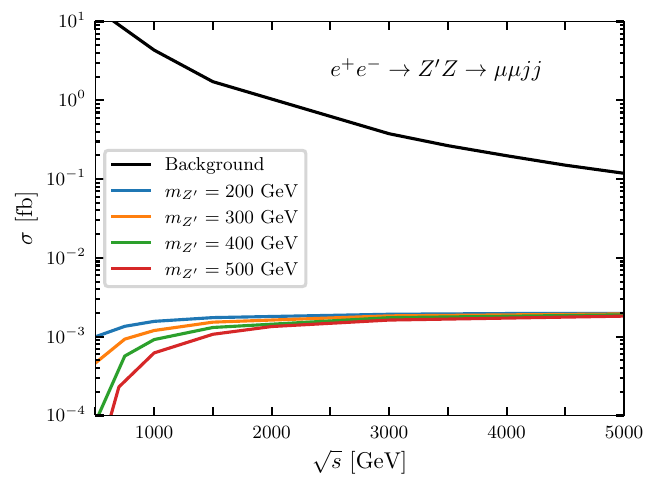}
\includegraphics[width=0.495\linewidth]{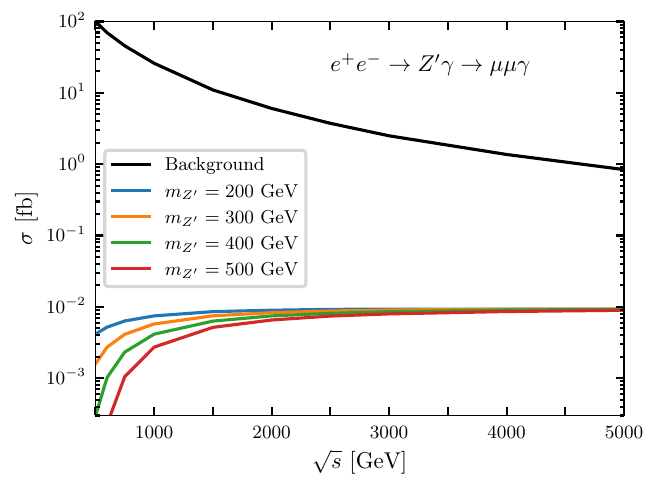}
\caption{ \label{fig:xsec}{Signal and background cross section for the process $e^+ e^- \to Z^*/\gamma^* \to Z' Z$ (left) and $e^+ e^- \to Z^*/\gamma^* \to Z' \gamma$ (right) as function of $\sqrt{s}$, for different $m_{Z'}$ values.} }
\end{figure}

We consider the non-resonant production of $Z'Z$ and $Z'\gamma$ through the anomalous coupling of three gauge bosons, with $Z'$ decaying into muons and $Z$ decaying into jets to maximize the signal cross sections.
Fig.~\ref{fig:xsec} displays the cross sections for both the signals and background processes with the default set of cuts provided by \texttt{MadGraph}, considering different $Z'$ masses and couplings consistent with trident bounds. Note that for all considered $Z'$, the cross section vanishes at threshold and increases monotonically with the center-of-mass energy until it becomes nearly constant. This behavior, which can be understood from Eq.~(\ref{fotonZL}) and Eq.~(\ref{ZetaZL}), arises from the anomaly and its apparent violation of Unitarity. 
An interesting aspect already mentioned at the end of Sect.~\ref{sec:muoncoll} is that the rise to a constant cross section in this case happens at a much smaller value of $\sqrt{s}$ than for the case in which the $Z'$ was as an intermediate state in the propagator, compare Fig.~\ref{fig:xsec_ZZ} with Fig.~\ref{fig:xsec} \footnote{This earlier reach in $\sqrt{s}$ to a constant cross section provides a further reassurance that the effective theory is under control at the energies considered since $\sqrt{s}\ll \Lambda$.}.   In contrast, the main SM backgrounds decrease with $\sqrt{s}$. Hence, to improve the signal-to-background ratio, higher center-of-mass energies are advantageous. For this reason, we focus on the CLIC collider, which is projected to reach $\sqrt{s} = 3 $ TeV with an integrated luminosity of up to $ 5 $ ab$^{-1}$\cite{CLIC:2018fvx, Aicheler:2018arh} ($ 5 $ ab$^{-1}$ in $7$ years with $708$ fb$^{-1}/$year). Notice that the asymptotic value of the signal cross section is larger for $Z'\gamma$ than for the case of $Z'Z$ in the final state by roughly an order of magnitude. However, the background is also roughly an order of magnitude larger.

\subsubsection{$e^+ e^- \to Z^*/\gamma^* \to Z' Z$}

At high energies, the cross section does not seem to vary much with the $Z'$ mass. However, the background cross section is two orders of magnitude larger than the signal, making it necessary to apply additional cuts to enhance the signal-to-background ratio. Specifically, we impose a window on the invariant mass of the final-state muon pair, $|m_{\mu^+ \mu^-}-m_{Z'}| < 10 $ GeV \footnote{The decay width of the $Z'$ increases as $\Gamma_{Z'} \propto g_{\mu}^2 m_{Z'} \propto m_{Z'}^3$ (see Eq.~\eqref{eq:trident}). This causes the $Z'$ resonance to broaden as $m_{Z'}$ increases, leaving less signal within a 10 GeV window, which results in a lower significance at higher masses. Another possibility would be to use an invariant mass window with a width proportional to $\Gamma_{Z'}$ to avoid losing signal events, but in that case, less background is removed. Ultimately, the significance does not improve significantly with this new window, so we chose a fixed-width window, which is easier to implement. }, as well as the angular cut $|\eta_{\mu^+ \mu^-}| < 1$, which takes advantage of the distinct angular distributions of the signal and background. Finally, we apply a cut on the invariant mass of the jet system, $40 $ GeV $< m_{jets} < 110 $ GeV.

The applied cuts are:

\begin{itemize}

\item Selection cuts: at least 1 jet with $p_{Tj} > 20 $ GeV, and at least 1 $\mu^+$ and 1 $\mu^-$ with $p_{T \mu^\pm} > 10 $ GeV.

\item At least a pair of muons with $|m_{\mu^+ \mu^-}-m_{Z'}|<10 $ GeV and $|\eta_{\mu^+ \mu^-}|<1$.

\item Invariant mass of the sum of jets (hadronic) $40$ GeV $<m_{jets}< 110$ GeV.

\end{itemize}

As an example, we provide the cutflows for the mass of $m_{Z'} = 200$ GeV in the Table \ref{tab:tabla_200_3tev_2} for the dedicated search of the $\mu \mu j j$ signal in CLIC. The total integrated luminosity is set as $5$ ab$^{-1}$.

\begin{table}
\begin{center}
\begin{tabular}{l c c }
\hline 
\hline 
 & Signal & Background \\ 
\hline 
Generator level cuts & 9.55 & 1022.4 \\ 
\hline 
Selection cuts & 8.79 & 901.28 \\ 
$|m_{\mu^+ \mu^-}-m_{Z'}| < 10 $ GeV & 8.01 & 12.03 \\ 
$40 $ GeV $< m_{jets} < 110 $ GeV & 7.22 & 6.95 \\ 
$|\eta_{\mu^+ \mu^-}| < 1$ & 5.18 & 0.265 \\ 
\hline 
\hline 
\end{tabular}
 \end{center}
\caption{\label{tab:tabla_200_3tev_2}Cutflow for the number of signal and background events in the $e^+ e^- \to Z' Z \to \mu \mu j j$ channel at $3~{\rm TeV}$ at CLIC. Events are normalized using the estimated cross sections with a total integrated luminosity of $\mathcal{L}=5~{\rm ab}^{-1}$. Signal is simulated with $m_{Z'}=200~{\rm GeV}$ and $g_\mu=0.445$. The selection cuts are described in the text.}
\end{table}

\begin{figure}[h]
\center
\includegraphics[width=0.7\linewidth]{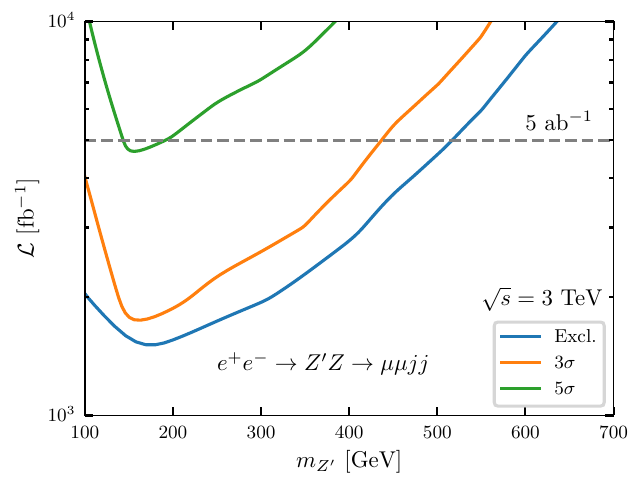}
\caption{ \label{fig:3tev_2}{Integrated luminosity required for exclusion, $3 \sigma$ and $5 \sigma$ significance, for the process $e^+ e^- \to Z^*/\gamma^* \to Z' Z$, with $\sqrt{s}= 3$ TeV. The dashed black line represents the estimated maximal luminosity CLIC will achieve.} }
\end{figure}

In Fig.~\ref{fig:3tev_2}, we show the discovery and exclusion regions as a function of $m_{Z'}$ based on the cuts mentioned earlier. The sharp increase in necessary luminosity at $m_{Z'}\lesssim 150$ GeV is due to the loss in efficiency in the $|m_{\mu^+ \mu^-}-m_{Z'}|<10 $ GeV cut from the $e^{+}e^{-}\to Zjj$ background. Note that discovery at $5\sigma$ is  attainable with CLIC  at its highest projected luminosity for $m_{Z'}\in [125,200]$ GeV and $3\sigma$ evidence can be obtained up to $m_{Z'}\lesssim 430$ GeV. Moreover, the mass range $m_{Z'}\in [100, 500]$ GeV could also be excluded at 95 $\%$ C.L. In a sense $e^{+}e^{-}$ collider non-resonant searches for triple-gauge anomalous couplings within our model are complementary to resonant $\mu^{+}\mu^{-}$ collider searches, since both colliders turn out to be able to probe different ranges of $Z'$ masses at $3\sigma$, with $e^{+}e^{-}$ in the $100\; {\rm GeV}\lesssim m_{Z'}\lesssim 430$ GeV while $\mu^{+}\mu^{-}$  probing the range $m_{Z'}\in [380, 700]$ GeV, the latter as can be seen in Fig.~\ref{fig:lumi_ZZ_4j}.

\subsubsection{$e^+ e^- \to Z^*/\gamma^* \to Z' \gamma$}

In addition to the cuts on the $Z'$ mass window and the angular distribution, we require the final photon to have a transverse momentum greater than 1.2 TeV, which we find to be the optimal cut. This cut leverages the fact that, for the signal, the photon should carry half of the collision energy since it is a two-to-two process, whereas this is not the case for the main Drell-Yan background.

The applied cuts are:

\begin{itemize}

\item Selection cuts: at least 1 photon with $p_{T \gamma} > 20 $ GeV, at least 1 $\mu^+$ and 1 $\mu^-$ with $p_{T \mu^\pm} > 10 $ GeV.

\item At least one pair of muons with $|m_{\mu^+ \mu^-}-m_{Z'}|<10 $ GeV and $|\eta_{\mu^+ \mu^-}|<1$.

\item Transverse momentum of the leading photon $p_{T \gamma} > 1.2 $ TeV.

\end{itemize}

As an example, we provide the cutflows for the mass of $m_{Z'} = 200$ GeV in the Table \ref{tab:tabla_200_mumua} for the dedicated search of the $\mu \mu \gamma$ signal at CLIC. The total integrated luminosity is set as $5$ ab$^{-1}$.

\begin{table}
\begin{center}
\begin{tabular}{l c c }
\hline 
\hline 
 & Signal & Background \\ 
\hline 
Generator level cuts & 46.49 & 12841 \\ 
\hline 
Selection cuts & 39.73 & 10253.2 \\ 
$|m_{\mu^+ \mu^-}-m_{Z'}| < 10 $ GeV & 36.05 & 79.4 \\ 
$|\eta_{\mu^+ \mu^-}| < 1$ & 25.78 & 18.2 \\ 
$p_{T \gamma} > 1200$ GeV & 19.28 & 8.8 \\ 
\hline 
\hline 
\end{tabular}
 \end{center}
\caption{\label{tab:tabla_200_mumua}Cutflow for the number of signal and background events in the $e^+ e^- \to Z' \gamma \to \mu \mu \gamma$ channel at $3~{\rm TeV}$ at CLIC. Events are normalized using the estimated cross sections with a total integrated luminosity of $\mathcal{L}=5~{\rm ab}^{-1}$. Signal is simulated with $m_{Z'}=200~{\rm GeV}$ and $g_\mu=0.445$. The selection cuts are described in the text.}
\end{table}

\begin{figure}[h]
\center
\includegraphics[width=0.7\linewidth]{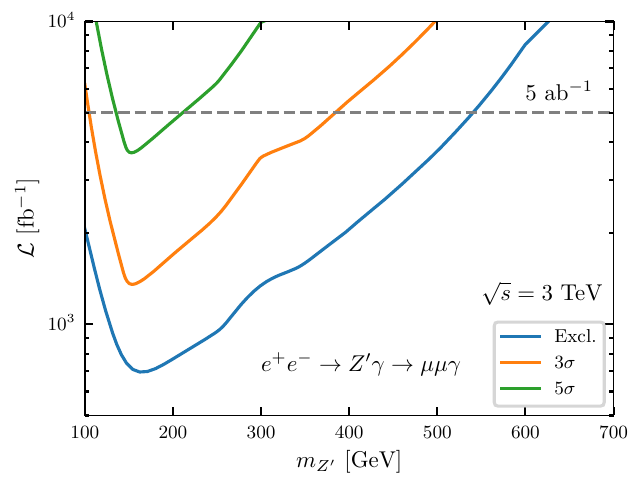}
\caption{ \label{fig:3tev_2_mumua}{Integrated luminosity required for exclusion, $3 \sigma$ and $5 \sigma$ significance, for the process $e^+ e^- \to Z^*/\gamma^* \to Z' \gamma$, with $\sqrt{s}= 3$ TeV. The dashed black line represents the estimated maximal luminosity CLIC will achieve.} }
\end{figure}

In Fig.~\ref{fig:3tev_2_mumua}, we present the discovery and exclusion regions as a function of $m_{Z'}$ based on the cuts mentioned earlier.  Once again a sharp increase in the necessary luminosity is observed for $m_{Z'}\lesssim 150$ GeV due to the loss in efficiency in the $Z'$ mass window cut from the $e^{+}e^{-}\to Z\gamma$ background. We see that there exists the possibility of discovery at CLIC for $m_{Z'}\in [125,225]$ GeV. Interestingly, there is an abrupt change in the significance for $m_{Z'}\approx  300 \sim 350$ GeV that can be traced back to the requirement on the photon transverse momenta, $p_{T \gamma}$. There are two main sources for the SM background, $e^{+}e^{-}\to Z\gamma$ and $e^{+}e^{-}\to \mu^{+}\mu^{-}\gamma$. We checked that imposing the $p_{T \gamma}>1.2$ TeV cut before the di-muon invariant mass cut, leads to a strong suppression for the $e^{+}e^{-}\to \mu^{+}\mu^{-}\gamma$ background,  shifting its peak in the di-muon invariant mass to values of order $m_{\mu^{+}\mu^{-}}\approx  300 \sim 350$ GeV, thus leaking more background once the $|m_{\mu^+ \mu^-}-m_{Z'}|<10 $ GeV cut is imposed for $m_{Z'}\in [300,350]$ GeV. This explains the shifts in the significances that are appreciated in Fig.~\ref{fig:3tev_2_mumua}. Evidence ($3\sigma$) for the anomalous triple gauge couplings can be obtained in the range $100\; {\rm GeV} \lesssim m_{Z'}\lesssim 400$ GeV, slightly smaller than in the $Z'Z$ final state case. Exclusions for the maximum coupling values allowed by trident could be achieved for $m_{Z'}\in[100,540]$ GeV.

\section{Conclusion}
\label{sec:conclusion}
We have studied the capabilities of current and future hadron and lepton ($e^{+}e^{-}$ and $\mu^{+}\mu^{-}$) colliders at probing triple-gauge couplings from mixed quantum gauge anomalies in an Abelian $U(1)'_\mu$ EFT extension of the SM model, under which second generation leptons are charged. In the EFT besides the SM particle content, we also have the associated $Z'$ from the spontaneous breaking of the $U(1)'_\mu$ and the non-decoupled gauge anomalous couplings involving the longitudinal $Z'$ polarization. The latter, loop-induced in nature, can potentially lead to non-Unitarity behaviors in cross sections in some energy ranges.  

Focusing in particular on the $Z'ZZ$, $Z'Z\gamma$ and $Z'\gamma\gamma$ couplings with the largest possible value allowed by neutrino trident (tree-level constraint), we find that the LHC in its high luminosity version cannot probe them due to the small signals provided in comparison to the large SM backgrounds.
The situation greatly improves for the $100$ TeV collider at its maximum considered luminosity of $\mathcal{L}=20~\mathrm{ab}^{-1}$ for the $Z'\gamma$ channel in particular. Evidence can be obtain for $m_{Z'}\in[150,800]~\mathrm{GeV}$ and even discovery for $m_{Z'}\in[230,330]~\mathrm{GeV}$. Lepton colliders and particularly, a muon collider resonantly producing the $Z'$ with 1 ab$^{-1}$ of integrated luminosity and an $e^{+}e^{-}$ collider such as CLIC at its highest projected luminosity of 5 ab$^{-1}$ producing the $Z'$ in association with a photon or a $Z$, are able to exclude $Z'$ masses in the range $m_{Z'}\in [280, 850]$ GeV for the former and  $m_{Z'}\in [100, 540]$ GeV for the latter. 
Evidence could be approximately achieved for $m_{Z'}\in [380, 700]$ GeV for the resonant muon collider, whereas for CLIC evidence could be found for $m_{Z'}\in [100, 430]$ GeV.
Interestingly enough, 5$\sigma$ discovery seems possible at CLIC in a range of $Z'$ masses $m_{Z'}\in [125, 225]$ GeV (roughly), the exact numbers depending on the anomalous coupling. These results suggest that the sensitivities of the considered lepton colliders are complementary, with the muon collider being more suited to explore larger $m_{Z'}$ values and the $e^+e^-$ collider allowing to probe masses as small as 100 GeV. Finally, we would like to stress that it is in the $Z'$ production at an $e^{+}e^{-}$  with $\sqrt{s}\gg m_{Z'}$ that the non-Unitary nature of the the triple gauge couplings allows to exploit the constant behavior of the signal cross sections in contrast with the suppressed behavior of the anomaly-free SM background.

    \section*{Acknowledgments}
This work has been partially supported by CONICET and by ANPCyT via project PICT-2018-03682. C. Wagner is supported by the U.S. Department of
Energy under contracts No. DEAC02- 06CH11357 at Argonne National Laboratory. The work of C. Wagner at the
University of Chicago has been supported by the DOE grant DE-SC0013642. C. Wagner would like to thank Perimeter Institute for its hospitality during the final part of this work. G. Zapata would like to express his gratitude for the financial support from the Dirección de Fomento de la Investigación at Pontificia Universidad Católica del Perú, through Grant No. DFI-PUCP PI0758 as well as from the Vicerrectorado de Investigación at Pontificia Universidad Católica del Perú through the Estancias Posdoctorales en la PUCP 2023 program.

\appendix

\section{Triple gauge anomalous couplings}
\label{appendix:couplings}

In the anomalous EFT, the 3-point vertex function is composed of two linearly divergent loop diagrams, where the momentum integrated in one of the loops can be shifted with respect to the other.
Since the anomalies do not cancel out, the resulting Ward identities depend on this choice of momentum shift.
In particular, there is no momentum shift that fixes simultaneously all Ward identities to zero.

Let us consider the example 
 of a  $U(1)'_\mu\times U(1)^2$ anomaly with the $Z'$ coupling to two massless vectors at loop level, via a massless fermion with   
 $U(1)$ and $U(1)'$ charges $q,q'$, respectively, and a heavier spectator of mass $M$ and $U(1)$ and $U(1)'$ charges $Q,Q'$ respectively, running in the loop. The the resulting Ward identities are,
 \begin{align}
(p+k)_{\rho}A^{\rho\mu\nu} & =\left[ \frac{q^{2}q^{\prime}}{4\pi^{2}}(w-z) - \frac{Q^{2}Q^{\prime}}{\pi^{2}}M^{2}I_{0}(M,p,k)+\frac{Q^{2}Q^{\prime}}{4\pi^{2}}(w-z)\right]\epsilon^{\lambda\mu\nu\sigma}p_{\lambda}k_{\sigma} , \label{eq:WardUVZp}\\
p_{\mu}A^{\rho\mu\nu} & = -\left[\frac{q^{2}q^{\prime}}{4\pi^{2}}(w-1)+\frac{Q^{2}Q^{\prime}}{4\pi^{2}}(w-1)\right]\epsilon^{\lambda\rho\nu\sigma}p_{\lambda}k_{\sigma} , \\
k_{\nu}A^{\rho\mu\nu} & = -\left[\frac{q^{2}q^{\prime}}{4\pi^{2}}(z+1)+\frac{Q^{2}Q^{\prime}}{4\pi^{2}}(z+1)\right]\epsilon^{\lambda\rho\mu\sigma}p_{\lambda}k_{\sigma} ,
\end{align}
where the momentum integrated in the loop is $l^\mu=zp^\mu+wk^\mu$.
If the ``covariant anomaly'' choice is used, which corresponds to setting $w=-z=1$, the Ward identities for the $U(1)$ gauge bosons are satisfied, both in the UV and after decoupling the spectator fermions.
The Ward identity for $Z'$ is then
\begin{equation}
    (p+k)_{\rho}A^{\rho\mu\nu} =\left[ \frac{q^{2}q^{\prime}}{2\pi^{2}} - \frac{Q^{2}Q^{\prime}}{\pi^{2}}M^{2}I_{0}(M,p,k)+\frac{Q^{2}Q^{\prime}}{2\pi^{2}}\right]\epsilon^{\lambda\mu\nu\sigma}p_{\lambda}k_{\sigma} , 
\end{equation}
and in the limit of the heavy spectator decoupling, $M^2\gg p^2,k^2,p\cdot k$, for which we get $M^2I_0(M,p,k)\to1/2$, one obtains,
\begin{equation}
    (p+k)_{\rho}A^{\rho\mu\nu} = \frac{q^{2}q^{\prime}}{2\pi^{2}}\epsilon^{\lambda\mu\nu\sigma}p_{\lambda}k_{\sigma},
\end{equation}
with no dependence on the spectator charges, which shows that, with this choice of momentum shift, the Ward identities in the EFT are independent of the UV physics.
Other choices of $w,z$, such as the ``consistent anomaly" that makes the Ward identities symmetrical among the three legs, do not lead to this cancellation and add extra contributions to the Ward identity of the $Z'$, which are identified with Wess-Zumino-Witten  counterterms.
Requiring that the Ward identities for the SM bosons are satisfied in the EFT then  is equivalent to setting a specific WZW counterterm that cancels the shift-dependent terms in the Ward identities, and therefore there is a one-to-one correspondence between the shift and the coefficients of the counterterms.
 In particular, the covariant anomaly does not require such WZW  counterterms in the EFT, so they are set to zero in our calculations.

We show in the following the explicit form of our triple-gauge-boson vertices using the Rosenberg parametrization \cite{Rosenberg:1962pp, Racioppi:2009yxa, Anastasopoulos:2008jt}, which are necessary for the computation of amplitudes. Let us recall that the most general expression that is Lorentz covariant is given by the Rosenberg parameterization in Eq.~\eqref{rosenberg}.
The convergent form factors $A_3,\dots, A_6$ are directly calculated from the loop integrals of SM fermions.
The form factors $\tilde{A}_1$, $\tilde{A}_2$ contain the divergent part of the loop integrals which are sensitive to the choice of momentum shift.
These factors are completely fixed in our setup by imposing the Ward identities corresponding to the EW gauge bosons as in Eqs.~\eqref{eq:wardp}--\eqref{eq:wardq} and solving for $\tilde{A}_1$, $\tilde{A}_2$.

We then find that the $Z'\gamma\gamma$ vertex reads,
\begin{equation}
\begin{aligned}
A_{\rho \mu \nu}^{Z' \gamma\gamma} &= -\frac{1}{2\pi^2} g' e^2 \left( 
\tilde{A}^{\gamma\gamma}_1\, \epsilon_{\alpha \mu \nu \rho}\,p^\alpha
+ \tilde{A}^{\gamma\gamma}_2\, \epsilon_{\alpha \mu \nu \rho}\,q^\alpha 
+ {A}^{\gamma\gamma}_3 \,\epsilon_{\alpha \beta \mu \rho} \, p^\alpha q^\beta p_{\nu}  \right. \\
& + {A}^{\gamma\gamma}_4\, \epsilon_{\alpha \beta \mu \rho} \, p^\alpha q^\beta q_{\nu} 
+ {A}^{\gamma\gamma}_5\, \epsilon_{\alpha \beta \nu \rho} \, p^\alpha q^\beta p_{\mu} 
+ {A}^{\gamma\gamma}_6\, \epsilon_{\alpha \beta \nu \rho} \, p^\alpha q^\beta q_{\mu}\Big) \;, \label{eq:vtxZpAA}
\end{aligned}
\end{equation}
with coefficients
\begin{eqnarray}
    A_i^{\gamma\gamma} &=& 2 Q_\mu ~I_i(p,q,m_\mu)\;,~~i=3,\dots,6\\
    \tilde{A}_1^{\gamma\gamma} &=& q^2 A_4^{\gamma\gamma} + p\cdot q~ A_3^{\gamma\gamma}\\
    \tilde{A}_2^{\gamma\gamma} &=& p^2 A_5^{\gamma\gamma} + p\cdot q~ A_6^{\gamma\gamma}
\end{eqnarray}
and vertex integrals
\begin{eqnarray}
    I_3(p,q,m_f) &=& \int_0^1 dx \int_0^{1-x}dy \frac{-xy}{y(1-y)p^2+x(1-x)q^2+2xy(p\cdot q)-m_f^2} \\
    I_5(p,q,m_f) &=& \int_0^1 dx \int_0^{1-x}dy \frac{-y(y-1)}{y(1-y)p^2+x(1-x)q^2+2xy(p\cdot q)-m_f^2} \\
    I_4(p,q,m_f) &=& -I_5(q,p,m_f) \\
    I_6(p,q,m_f) &=& -I_3(p,q.m_f)    
\end{eqnarray}

The $Z'Z\gamma$ vertex reads,
\begin{equation}
\begin{aligned}
A_{\rho \mu \nu}^{Z' Z\gamma} &= -\frac{1}{4\pi^2} g' g_Z e \left( 
\tilde{A}^{Z\gamma}_1\, \epsilon_{\alpha \mu \nu \rho}\,p^\alpha
+ \tilde{A}^{Z\gamma}_2\, \epsilon_{\alpha \mu \nu \rho}\,q^\alpha 
+ {A}^{Z\gamma}_3 \,\epsilon_{\alpha \beta \mu \rho} \, p^\alpha q^\beta p_{\nu}  \right. \\
& + {A}^{Z\gamma}_4\, \epsilon_{\alpha \beta \mu \rho} \, p^\alpha q^\beta q_{\nu} 
+ {A}^{Z\gamma}_5\, \epsilon_{\alpha \beta \nu \rho} \, p^\alpha q^\beta p_{\mu} 
+ {A}^{Z\gamma}_6\, \epsilon_{\alpha \beta \nu \rho} \, p^\alpha q^\beta q_{\mu}\Big) \;, \label{eq:vtxZpZA}
\end{aligned}
\end{equation}
where the coefficients are
\begin{eqnarray}
    A_i^{Z\gamma} &=& -2 Q_\mu \left(-\frac{1}{2} + 2 s_W^2\right)~I_i(p,q,m_\mu)\;,~~i=3,\dots,6\\
    \tilde{A}_1^{Z\gamma} &=& q^2 A_4^{Z\gamma} + p\cdot q ~ A_3^{Z\gamma}\\
    \tilde{A}_2^{Z\gamma} &=& p^2 A_5^{Z\gamma} + p\cdot q ~ A_6^{Z\gamma}
\end{eqnarray}
The $Z$ goldstone contribution vanishes since the corresponding vertex $A^{Z'G\gamma}_{\rho\nu}$ includes only muons, which have vector-like couplings to $\gamma$, and axial couplings to $G$ and $Z'$, and therefore the loop integral vanishes.

The $Z'ZZ$ vertex reads,
\begin{equation}
\begin{aligned}
A_{\rho \mu \nu}^{Z' ZZ} &= -\frac{1}{8\pi^2} g' g_Z^2 \left( 
\tilde{A}^{ZZ}_1\, \epsilon_{\alpha \mu \nu \rho}\,p^\alpha
+ \tilde{A}^{ZZ}_2\, \epsilon_{\alpha \mu \nu \rho}\,q^\alpha 
+ {A}^{ZZ}_3 \,\epsilon_{\alpha \beta \mu \rho} \, p^\alpha q^\beta p_{\nu}  \right. \\
& + {A}^{ZZ}_4\, \epsilon_{\alpha \beta \mu \rho} \, p^\alpha q^\beta q_{\nu} 
+ {A}^{ZZ}_5\, \epsilon_{\alpha \beta \nu \rho} \, p^\alpha q^\beta p_{\mu} 
+ {A}^{ZZ}_6\, \epsilon_{\alpha \beta \nu \rho} \, p^\alpha q^\beta q_{\mu}\Big) \;, \label{eq:vtxZpZZ}
\end{aligned}
\end{equation}
and the coefficients are
\begin{eqnarray}
    A_i^{ZZ} &=& \left[2 Q_\mu \left(-\frac{1}{2} + 2 s_W^2\right)^2 + \frac{Q_\mu}{2}\right]~I_i(p,q,m_\mu)+ Q_\mu ~ I_i(p,q,0) \;,~~i=3,\dots,6\\
    \tilde{A}_1^{ZZ} &=& q^2 A_4^{ZZ} + p\cdot q ~ A_3^{ZZ} - \frac{1}{3} \frac{Q_\mu}{2}m_\mu^2 ~I_0(p,q,m_\mu)\\
    \tilde{A}_2^{ZZ} &=& p^2 A_5^{ZZ} + p\cdot q ~ A_6^{ZZ} + \frac{1}{3} \frac{Q_\mu}{2}m_\mu^2 ~I_0(p,q,m_\mu)
\end{eqnarray}
where the integral in the goldstone contribution is
\begin{equation} 
    I_0(p,q,m_f) = \int_0^1 dx \int_0^{1-x}dy \frac{-1}{y(1-y)p^2+x(1-x)q^2+2xy(p\cdot q)-m_f^2}
\end{equation}

\bibliographystyle{biblio/bibstyle}
\bibliography{biblio/biblio.bib}

\end{document}